\documentclass[preprintnumbers,showpacs,prd,aps,floatfix,preprint,nofootinbib,superscriptaddress,dvipdfmx]{revtex4}
\usepackage[dvipdfmx]{graphicx,color}
\usepackage{epstopdf}
\usepackage{amssymb}
\usepackage{amsmath}
\usepackage{subfigure}
\usepackage{epsfig}
\usepackage{float}
\usepackage{epsf}
\usepackage{latexsym}

\makeatletter

\newcommand{\figcaption}[1]{\def\@captype{figure}\caption{#1}}
\newcommand{\tblcaption}[1]{\def\@captype{table}\caption{#1}}
\newcommand{\Nf}{N_f}
\newcommand{\kl}{\kappa_l}
\newcommand{\kh}{\kappa_h}
\newcommand{\khc}{\kappa_{h_c}}

\newcommand{\ml}{m_l}
\newcommand{\no}{\nonumber}
\makeatother

\def\simge{\mathrel{%
       \rlap{\raise 0.511ex \hbox{$>$}}{\lower 0.511ex \hbox{$\sim$}}}}
\def\simle{\mathrel{
       \rlap{\raise 0.511ex \hbox{$<$}}{\lower 0.511ex \hbox{$\sim$}}}}

\begin{document}

\title{
Exploring the nature of chiral phase transition in two-flavor QCD using
extra heavy quarks
}
\author{Shinji~Ejiri}
\affiliation{
 Department of Physics, Niigata University, Niigata 950-2181, Japan}
\author{Ryo~Iwami}
\affiliation{
 Graduate School of Science and Technology, Niigata University,
 Niigata 950-2181, Japan}
\author{Norikazu~Yamada}
\affiliation{
KEK Theory Center, Institute of Particle and Nuclear Studies, High
Energy Accelerator Research Organization (KEK), Tsukuba 305-0801,
Japan}
\affiliation{
School of High Energy Accelerator Science,
SOKENDAI (The Graduate University for Advanced Studies),
Tsukuba 305-0801, Japan}

\date{\today}

\begin{abstract}
 Chiral phase transition of two-flavor QCD at finite quark masses
 is known to be a crossover except near the chiral limit, but it can turn
 to a first order transition when adding many extra flavors.
 This property is used to explore the nature of the phase transition of
 massless two-flavor QCD using lattice numerical simulations.
 The extra heavy flavors being incorporated in the form of the hopping
 parameter expansion through the reweighting, the number of the extra
 flavors and their masses appear only in a single  parameter, defined by
 $h$.
 We determine the critical value of $h$, at which the
 first order and the crossover regions are separated, and examine its
 dependence on the two-flavor mass.
 The lattice calculations are carried out at $N_t=4$, and show that the
 critical value of $h$ does not depend on the two-flavor mass in the
 range we have studied ($0.46 \le m_\pi/m_\rho \le 0.66$)
 and appears to remain finite and positive in the chiral limit,
 suggesting that the phase transition of massless two-flavor QCD is of
 second order.
\end{abstract}


\maketitle

\section{Introduction}
 \label{sec:introduction}

Quantum chromodynamics (QCD)
shows a variety of phases, and in passing over a phase
boundary one would encounter either first (discontinuous) or second
order (continuous) transition, depending on temperature, density, quark
masses, the number of flavors, etc.
Chiral phase transition of QCD with two massless quarks at the
vanishing chemical potential has been studied with various approaches
for a long time, since it provides us with a solid basis in the study
of $2+1$-flavor QCD in the real world.
Nevertheless, the nature of the transition of this relatively simple
system is yet ambiguous, and is counted as one of the longstanding
problems.

Based on the universality argument and the results of the leading
order $\epsilon$ expansion, Pisarski and Wilczek analyzed the
renormalization group (RG) flow of the three-dimensional scalar field
theory, which shares the same internal symmetry with massless QCD around
the critical temperature ($T_c$), and pointed out that, in the
two-flavor case, the order of the transition could crucially depend on
the presence (or the absence) of the flavor singlet axial ($U_A(1)$)
symmetry at $T_c$~\cite{Pisarski:1983ms}.
If $U_A(1)$ symmetry is largely violated at $T_c$, the second order
phase transition with the $O(4)$ scaling becomes possible although not
mandatory.
On the other hand, when the symmetry is effectively and fully restored at
$T_c$, the leading order calculation of the $\epsilon$ expansion
suggests no infrared fixed point (IRFP), and hence the second order
phase transition is excluded.
But, later, further studies using different advanced techniques
found evidence of IRFP, and the confirmation of the presence of the
IRFP is under active
investigation~\cite{Butti:2003nu,Aoki:2012yj,Pelissetto:2013hqa,Nakayama:2014sba}.
Thus, the transition in this case again can be either of first or
second order.
Recently, a novel possibility is pointed out, following the RG flow
analysis: in the presence of small but finite $U_A(1)$ symmetry
breaking, the system may undergo the second order transition with the
$O(4)$ scaling but one of the critical exponents related to the
scaling dimension of the leading irrelevant operator is different from
that of the $O(4)$~\cite{Sato:2014axa}, although again the second order
transition is not mandatory.

Numerical simulations of QCD on the lattice can, in principle,
determine the order of the transition as well as the universality class,
to which massless two-flavor QCD belongs, by performing the scaling
study~\cite{Karsch:1993tv,Karsch:1994hm,Iwasaki:1996ya,AliKhan:2000iz,
D'Elia:2005bv,Bonati:2009yg,RBCBi09,Kanaya:2010qj,Bonati:2014kpa}.
However, it is not easy to keep all the systematic and statistical
uncertainties under control in the chiral limit due to large
computational costs.
Furthermore, it appears that, in practice, the standard scaling study
may not be efficient enough to distinguish the first and the second
order transitions because the scaling functions are similar between the
$Z_2$ and $O(4)$ universality classes~\cite{Brandt:2013mba}.
With lattice QCD simulations, one can also study the presence (or
absence) of $U_A(1)$ symmetry through the Dirac spectrum.
For recent progress, see, for example,
Refs.~\cite{Cossu:2013uua,Buchoff:2013nra,Bhattacharya:2014ara,Dick:2015twa}.

Clarifying this point is important not only for understanding the QCD phase
diagram but also for the scenario of the axion dark matter.
The axion abundance is essentially determined by the temperature
dependence of the topological susceptibility, $\chi_t(T)$, which
vanishes when $U_A(1)$ symmetry is fully and effectively restored.
If $\chi_t$ vanishes very rapidly right above $T_c$, too many axions
would be produced, and the axion dark matter scenario becomes hard or is
even excluded~\cite{Kitano:2015fla}, depending on how rapidly it
vanishes.
The lattice studies to test the axion dark matter scenario has recently
begun in the quenched
approximation~\cite{Kitano:2015fla,Berkowitz:2015aua,Borsanyi:2015cka}.

In this paper, we follow the approach proposed in
Ref.~\cite{Ejiri:2012rr}, in which the phase transition of two-flavor
QCD is studied by adding many extra heavy quarks.
We call this the many flavor approach.
The transition of two-flavor QCD at a finite quark mass is known to be
a crossover, but it can turn to a first order transition when adding
many extra flavors.
This property is used to explore the nature of the phase transition of
massless two-flavor QCD.
The extra quarks are incorporated in the form of the hopping parameter
expansion (HPE) through the reweighting.
Then the number of the extra flavors ($N_f$) and their mass parameter
($\kh$) appear in a single parameter
\begin{eqnarray}
     h
 &=& 2\,N_f\,(2\kh)^{N_t} \ ,
  \label{eq:hcdef}
\end{eqnarray}
where $N_t$ denotes the number of lattice sites in the temporal
direction.
We determine the critical value of the parameter ($h_c$) at which the
first order and the crossover regions are separated, and examine its
dependence on the two-flavor mass.
The order of the transition for a given $h$ is discriminated by the
shape of the constraint effective potential at $T_c$, which is
constructed from the probability distribution function (PDF) for a
generalized plaquette.
Namely, it is discriminated by whether the potential at $T_c$ is in
single- or double-well shape.
It is important to note that, in the determination of $h_c$ in this
approach, the convergence of the HPE is not the matter since we can
consider arbitrary small $\kh$ by considering arbitrary large $\Nf$ as
seen from Eq.(\ref{eq:hcdef}).

We perform an exploratory study on $N_t=4$ and try to see how $h_c$
depends on the two-flavor mass.
Then, $h_c$ is found to stay constant against the change of the
two-flavor mass in the range we have studied and to remain positive and
finite in the chiral limit.
Since the two-flavor system is equivalent to the $2+N_f$-flavor system
with $h=0$, our result suggests that massless two-flavor QCD belongs to
the region of second order phase transition.
This kind of extension of QCD is useful also for the study of the phase
structure in the presence of finite chemical
potential~\cite{Ejiri:2012rr}.
The similar approach but introducing a finite imaginary chemical
potential is taken in Ref.~\cite{Bonati:2014kpa}.

The paper is organized as follows.
After the central idea of many flavor approach is explained in
Sec.~\ref{sec:approach}, the method is described in detail in
Sec.~\ref{sec:method}.
The lattice setup and the main part of this paper are given in
Sec.~\ref{sec:numerical_results}.
In Sec.~\ref{sec:consistency-check}, two independent analyses are
performed for the consistency check.
Finally, the conclusion and perspectives are stated in
Sec.~\ref{sec:summary}.
The preliminary result of this work is available in
Ref.~\cite{Ejiri:2015pva}.

\section{Many flavor approach}
\label{sec:approach}

\begin{figure}[tb]
\begin{center}
\begin{tabular}{cc}
\includegraphics*[width=0.5 \textwidth,clip=true]
{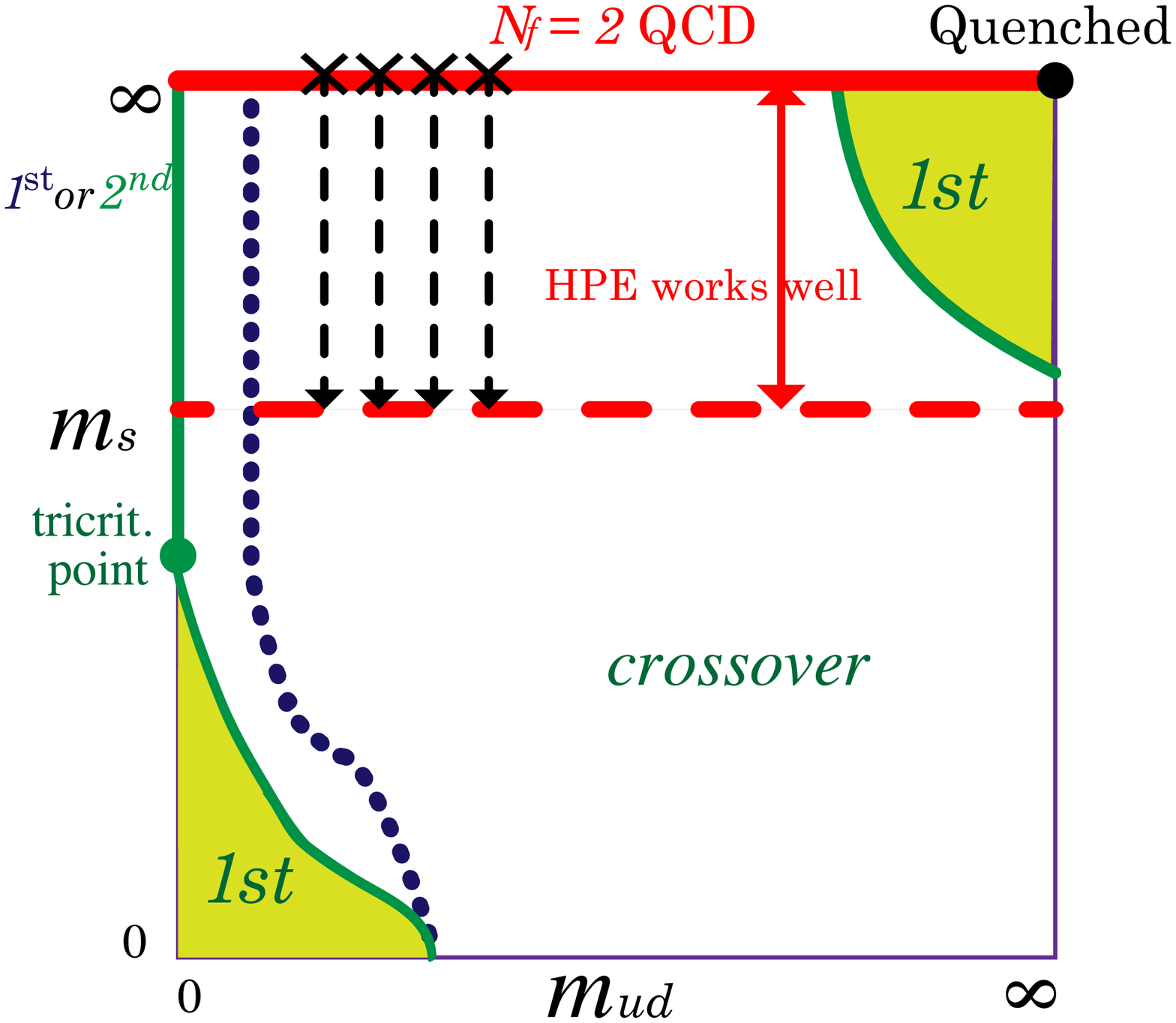}&
\includegraphics*[width=0.5 \textwidth,clip=true]
{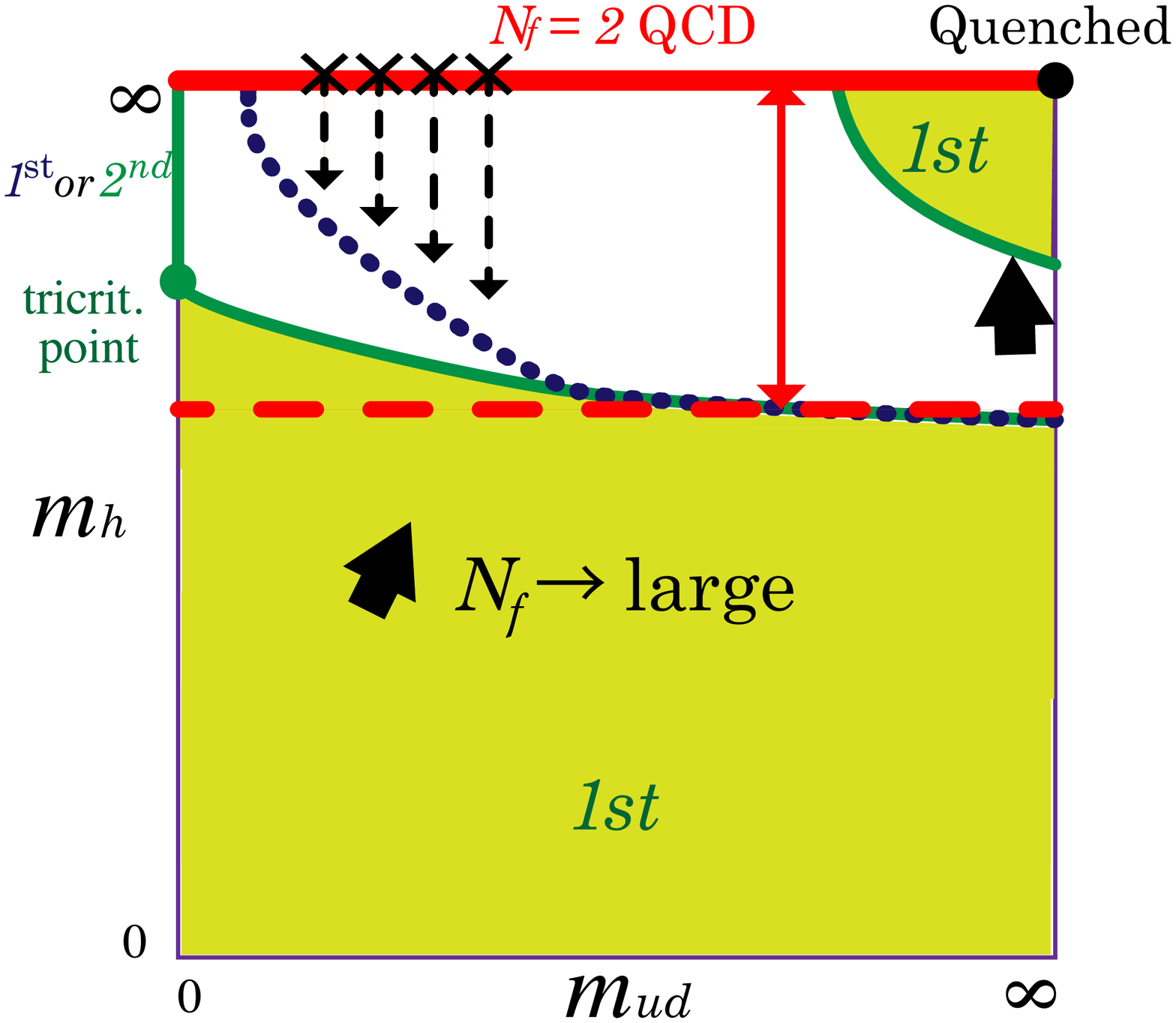}\\
 (a)\ 2+1 flavor QCD&
 (b)\ 2+$\Nf$ flavor QCD\\
\end{tabular}
\vspace{-1ex}
\caption{Basic idea of "many flavor approach".}
\label{fig:columbia-plot}
\end{center}
\end{figure}

The central idea of many flavor approach is outlined.
Figure~\ref{fig:columbia-plot} (a) shows the so-called Columbia plot for
$2+1$-flavor QCD~\cite{Brown:1990ev}, which summarizes the present
knowledge on the mass dependent nature of the phase transition of
QCD as a function of $m_{ud}$ and $m_s$.
The physical point is believed to be located in the crossover
region~\cite{RBCBi09,Aoki:2006we}.
The plot tells us that there are two distinct first order regions lying
around the quenched limit ($m_{ud}=m_s=\infty$) and the chiral limit of
three-flavor QCD ($m_{ud}=m_s=0$), respectively.
In what follows, we focus on the latter.

Our interest is in whether the massless two-flavor QCD point ($m_{ud}=0$
and $m_s=\infty$) is inside the first order region or not and hence in
the shape of the first order region.
If it ends at a finite $m_s$, called the tricritical point, massless
two-flavor QCD is on the line of the second order phase transition (the
solid curve on the $m_{ud}=0$ line).
On the other hand, if the first order region extends to $m_s=\infty$,
massless two-flavor QCD should undergo the first order transition ({\it
i.e.} the dotted curve near the $m_{ud}=0$ line).
In either case, if we could resolve which of the solid or dotted curves
is realized, we would be able to answer the question.
However, it is difficult to trace the critical line for the $2+1$-flavor
case, because the critical line is located in the small $m_s$ region
when $m_{ud}$ is moderately
small~\cite{RBCBi09,Ding:2013lfa,Jin:2014hea}.

The situation will change in $2+N_f$-flavor QCD as shown in
Fig.~\ref{fig:columbia-plot} (b).
The bottom-right corner of Fig.~\ref{fig:columbia-plot} (b) is of
first order for $N_f \geq 3$.
The critical line is expected to move upward as $\Nf$ increases, and for
sufficiently large $\Nf$ it could enter the region where the hopping
parameter expansion from the static limit (crosses at
$m_h=\infty$) works well (above the dashed line).
Then, one should be able to easily identify how the critical line runs
as a function of $m_{ud}$.
If the critical heavy mass $m_h^c(m_{ud})$ remains finite in the
$m_{ud}\to 0$ limit, it immediately means that massless two-flavor QCD
corresponding to the point $(m_{ud},m_h)=(0,\infty)$ is the outside of
the first order region.
An important remark is that, in the limit of $m_s\to\infty$ or
$m_h\to\infty$, both $2+1$- and $2+\Nf$-flavor QCD end up with the same
theory, which we want to study.
Thus, the original question is simplified to the one whether the
critical heavy mass in the chiral limit stays finite or not.

While most of the current knowledge on the shape of the first order
region are only qualitative, in a context of the Taylor-expanded
reweighting method~\cite{BS02} we can derive a solid statement about the
slope at $m_{ud}=m_h$ if the same quark action is employed for two- and
$\Nf$-flavors.
Suppose that $m_{ud}=m + \Delta m_{ud}$ and $m_{h}=m + \Delta m_h$ and
expand the logarithm of the quark determinants in terms of $\Delta
m_{ud}$ and $\Delta m_h$.
Then, one will be aware that the partition function does not change as
long as $\Delta m_h=- 2\, \Delta m_{ud}/\Nf$ and hence that physics is
identical along the line of slope $-2/\Nf$ near $m_{ud} = m_h$ line in
the $(m_{ud}, m_h)$ plane~\cite{ejiri04}.
This means that the critical line in $2+\Nf$-flavor QCD crosses the line
of $m_{ud}=m_h$ with a slope milder than the $2+1$ flavor case.

In Ref~\cite{Ejiri:2012rr}, it is demonstrated that by adding extra
flavors the end point (or the critical line) indeed enters the region
reachable by the hopping parameter expansion.
In this paper, we examine the light quark mass dependence of the
end point.
As we will explain below, $m_h^c(m_{ud})$ seems to remain finite in
the chiral limit of $m_{ud}$, suggesting the phase transition of
massless two-flavor QCD is of second order.

\section{Calculational method}
\label{sec:method}

We first generate two-flavor configurations at finite temperatures
following the standard hybrid Monte Carlo method.
Using the hopping parameter expansion and the reweighting method,
we incorporate extra $\Nf$ flavors of heavy quarks into those
configurations, and measure the probability distribution function
for a generalized plaquette to construct the constraint effective
potential for $2+\Nf$-flavor QCD.

After the HPE, the hopping parameter for heavy quarks $\kappa_h$ and
$N_f$ appear only in a single parameter, $h$ [see Eq.~(\ref{eq:h})].
If the parameter $h$ is in the crossover region, the effective potential
should take a single-well shape at the pseudocritical temperature,
$T_{pc}$.
On the other hand, when the parameter $h$ enters the first order region,
a double-well shape should emerge at $T_c$.
By scanning $h$, we determine the critical value $h_c$, at which the
first order and the crossover regions are separated.
The critical value $h_c$ is determined at four values of two-flavor
mass to see the light quark mass dependence of $h_c$.
In the following, the calculational procedure is described in detail.

The PDF was introduced in Refs.~\cite{Bruce:1981,Binder:1981sa} and has
been extensively used in various fields to study the critical
properties of various materials~\cite{Plascak:2013} or the phase
diagrams of QCD~\cite{whot11}.
In our study of 2+$\Nf$-flavor QCD, the PDF $w$ for a quantity $\hat X$
is defined by
\begin{eqnarray}
    w(X; \beta, \kl, \kh,\Nf)
&=& \int \!\! {\cal D} U\, \delta(X- \hat{X}) \ 
    \big[\det M(\kh)\big]^{\Nf}
    e^{-S_{\rm gauge}(\beta) - S_{\rm light}(\kl)},
\label{eq:pdist}
\end{eqnarray}
where $S_{\rm gauge}(\beta)$ and $S_{\rm light}(\kl)$ are the lattice
actions for the gauge field and two flavors of light quarks,
respectively.
$\beta=6/g_0^2$ is the simulation parameter setting the temperature
through the lattice spacing, and $\kl$ is the light quark mass
parameter.
Note that the method  described below works for any kinds of light quark
action unless it contains $\beta$ dependent coefficients.
The action for the $\Nf$ extra flavors is written in the determinant
form in Eq.~(\ref{eq:pdist}), where $M(\kh)$ is the lattice Dirac
operator for heavy quarks with a mass parameter $\kh$.
A quantity to be constrained, $\hat X$, is basically arbitrary, but the
order parameter would be the most natural choice if it is available.
In this paper, following previous
works~\cite{Ejiri:2007ga,whot11,Ejiri:2012rr,Ejiri:2013lia}, $\hat X$ is
chosen to be the generalized plaquette
\begin{eqnarray}
  \hat P
= c_0\, \hat W_P + 2 c_1\, \hat W_R\ ,
\end{eqnarray}
where $\hat W_P$ and $\hat W_R$ denote the averaged plaquette and
rectangle, respectively, and $c_0$ and $c_1$ satisfying $c_0=1-8 c_1$
are the improvement coefficients for lattice gauge action.
In terms of $\hat P$, the gauge action is written as
\begin{eqnarray}
    S_{\rm gauge}(\beta)
= - 6\, N_{\rm site}\, \beta\, \hat P\, ,
\end{eqnarray}
where $N_{\rm site}= N_{\rm s}^3 \times N_t$ represents the number of
sites in four-dimensional lattice volume.
Our main analysis is carried out with $\hat X=\hat P$, but the
calculation for $\hat X=\hat L$ with $\hat L$ the real part of the
Polyakov loop averaged over spatial sites is also performed as a
consistency check (see Sec.~\ref{subsec:polyakov}).
Choosing $\hat X=\hat P$ brings a great simplification in the
numerical analysis as explained below.
In principle, we could choose other quantities, {\it e.g.} the
chiral condensate, to be $\hat X$.
But, whenever a quantity other than $\hat P$ is chosen, we lose not only
the advantage for $\hat X=\hat P$ but also the accuracy in the results
as demonstrated in Sec.~\ref{subsec:polyakov} for $\hat X=\hat L$.

With the PDF thus obtained, the constraint effective potential $V$ is
calculated by
\begin{eqnarray}
       V(X;\beta,\kl,\kh,\Nf)
&=&  - \ln w(X;\beta,\kl,\kh,\Nf)
 \label{eq:effective-potential}
\end{eqnarray}
In practice, the PDF is not directly accessible, and hence we instead
calculate the histogram defined by
\begin{eqnarray}
    H(X;\beta,\kl,\kh,\Nf)
&=& \frac{w(X;\beta,\kl,\kh,\Nf)}{Z(\beta,\kl,\kh,\Nf)}\ ,
 \label{eq:histogram}
\end{eqnarray}
where $Z(\beta)$ is the partition function.
Note that $Z(\beta)$ is not calculable, but its ratio at two
different $\beta$ values is calculable~\cite{Ejiri:2008xt}.

In order to see the shape of the potential, we need to calculate the
potential over a certain range of $X$.
However, a simulation at a single $\beta$ provides the potential only in
a limited range of $X$.
Thus, the potential calculated at a certain $\beta$ needs to be
translated to that at other $\beta$.
We call the temperature, at which we want to calculate the potential,
the reference temperature (or $\beta_{\rm ref}$).
Then, the potential is calculated as
\begin{eqnarray}
&&      V(X;\beta_{\rm ref},\kl,\kh,\Nf) + \ln Z(\beta^*,\kl',\kh',\Nf')
\no\\
&=& - \ln H(X';\beta',\kl',\kh',\Nf')
    - \ln \left(\frac{w(X'';\beta'',\kl'' ,\kh'' ,\Nf'')}
                     {w(X';\beta'  ,\kl',\kh',\Nf')}
          \right)
    - \ln \left(\frac{w(X;\beta_{\rm ref},\kl ,\kh ,\Nf)}
                     {w(X'';\beta''       ,\kl'',\kh'',\Nf'')}
          \right)
\no\\
& & - \ln \frac{Z(\beta',\kl',\kh',\Nf')}{Z(\beta^*,\kl',\kh',\Nf')}
\ .
 \label{eq:effective-potential-2}
\end{eqnarray}
All the intermediate quantities such as $\beta'$, $X'$, $\kl'$, $\kh'$
and $\Nf'$ are arbitrary as well as $\beta^*$.
For a practical reason, we take
\begin{eqnarray}
 X'=X''=X,\ \ \
 \beta'' = \beta_{\rm ref},\ \ \
 \kl' = \kl'' = \kl,\ \ \
 \kh' = \kh'' = 0\,\ \ \
 \Nf' = \Nf'' = 0\ ,
 \label{eq:setup}
\end{eqnarray}
and $\beta^*$ is chosen to be the vicinity of $\beta_{\rm ref}$.

In the following, we take $\hat X=\hat P$, then the effective potential
is simplified as
\begin{eqnarray}
    V(P; \beta_{\rm ref},\kl,\kh,\Nf) + \ln Z(\beta^*,\kl',\kh',\Nf')
&=&   V_{\rm light}(P; \beta_{\rm ref},\kl)
    - \ln R(P;\beta_{\rm ref},\kl,\kh,\Nf)\ .
 \no\\
\label{eq:vefftrans}
\end{eqnarray}
The first term is defined by
\begin{eqnarray}
    V_{\rm light}(P; \beta_{\rm ref},\kl)
= - \ln H(P;\beta,\kl,0,0)  
  - 6\,N_{\rm site}\,(\beta_{\rm ref} - \beta)\,P
  - \ln \frac{Z(\beta,\kl,0,0)}{Z(\beta^*,\kl,0,0)}
\label{eq:V-two-flavor}
\end{eqnarray}
and represents the constraint effective potential for two flavors alone.
The second term of Eq.~(\ref{eq:vefftrans}) is defined by
\begin{eqnarray}
    R(P;\beta_{\rm ref},\kl,\kh,\Nf)
&=& \left\langle
    \displaystyle
    \left[ \det M(\kh) \right]^{\Nf}
    \right\rangle_{P: {\rm fixed},(\beta_{\rm ref},\kl)}, \ 
\label{eq:lnr}\\
    \langle \cdots \rangle_{P: {\rm fixed}, (\beta_{\rm ref},\kl)}
&\equiv&
    \frac{\langle \delta(P- \hat{P}) \cdots \rangle_{(\beta_{\rm ref},\kl)}}
         {\langle \delta(P- \hat{P}) \rangle_{(\beta_{\rm ref},\kl)}}\ ,
\label{eq:exp-fixedP}
\end{eqnarray}
where $\langle \cdots \rangle_{(\beta,\kl)}$ denotes the ensemble
average over two-flavor configurations generated with $\beta$ and $\kl$.
It is important to note that, separating the effective potential into
the two-flavor part and the extra heavy part as in
Eq.~(\ref{eq:vefftrans}), the latter becomes independent of
$\beta_{\rm ref}$.
The reason is as follows.
Due to the operator $\delta(P-\hat P)$, the factor of
$\exp(6 N_{\rm site}\beta_{\rm ref} P)$ comes out of the brackets in
Eq.~(\ref{eq:exp-fixedP}).
Since this factor cancels between the numerator and the denominator in
Eq.~(\ref{eq:exp-fixedP}), $\beta_{\rm ref}$ dependence disappears.
This simplification takes place only for $\hat X=\hat P$.

For the extra heavy quarks, we employ the unimproved Wilson fermion
because it suffices for the present purpose.
For sufficiently small $\kh$, the determinant in Eq.~(\ref{eq:lnr}) can
be approximated by the leading order of the HPE as
\begin{eqnarray}
   \ln \left[ \det M (\kh) \right]^{\Nf}
&\approx&
   \Nf
    \left( 288\, N_{\rm site}\, \kh^4\, \hat W_P
         + 12 N_s^3 (2 \kh)^4\, \hat{L}
    \right)
\no\\
&=& 6\,N_s^3\,h\,\hat Y
\label{eq:detmw-1}\\
&=&   9\,N_{\rm site}\,\frac{h}{c_0}\, \hat P
    + 6\,N_s^3\,h\,\hat Z \ ,
\label{eq:detmw-2}
\end{eqnarray}
where the following quantities have been introduced:
\begin{eqnarray}
    h
&=& 2\,\Nf\,(2 \kh)^4\ ,
\label{eq:h}\\
    \hat Y
&=& 6\,\hat W_P + \hat{L}\ ,
\label{eq:Y}\\
    \hat Z
&=& - \frac{12\,c_1}{c_0}\,\hat W_R + \hat{L}\ .
\label{eq:Z}
\end{eqnarray}
Here and hereafter, $N_t=4$ is assumed because we take that value in
numerical simulations, but the extension to other values of $N_t$ is
straightforward though the expression becomes complicated.
Then, $\ln R$ in Eq.~(\ref{eq:vefftrans}) can be approximated as
\begin{eqnarray}
    \ln R(P;\kl,h)
&\approx&
    \ln \left\langle
    \displaystyle
    \exp\left( 6\,N_s^3\,h\,\hat Y \right)
    \right\rangle_{P: {\rm fixed},(\beta,\kl)}
\label{eq:r-1}\\
&=&  9\,N_{\rm site}\,\frac{h}{c_0}\, P
   + \ln R'(P;\kl,h)
\label{eq:r-2}
\end{eqnarray}
where
\begin{eqnarray}
    \ln R'(P;\kl,h)
&=& \ln \left\langle
    \displaystyle
    \exp\left( 6\,N_s^3\,h\,\hat Z \right)
    \right\rangle_{P: {\rm fixed},(\beta,\kl)}\ .
\label{eq:rdash}
\end{eqnarray}
Although Eqs.~(\ref{eq:r-1}) and (\ref{eq:r-2}) are algebraically
identical, the equality is not necessarily trivial in numerical data
because the $\delta$ function is approximated by
\begin{eqnarray}
        \delta(x)
 \approx 1/(\Delta \sqrt{\pi}) \exp[-(x/\Delta)^2]\ .
 \label{eq:delta-func}
\end{eqnarray}
Then, the difference can arise when
\begin{eqnarray}
  \big\langle e^{6 N_s^3 h \hat Y}
          \exp\left[-(P-\hat P)^2/\Delta^2\right]
  \big\rangle_{(\beta,\kl)}
- e^{9 N_{\rm site} \frac{h}{c_0}P}
  \big\langle e^{6 N_s^3 h \hat Z}
  \exp\left[-(P-\hat P)^2/\Delta^2\right]
  \big\rangle_{(\beta,\kl)}
  \neq 0  ,\no
\end{eqnarray}
which should vanish for sufficiently small $\Delta$, but then the
statistical error will enlarge.
In the following analysis, both expressions are examined to check
the consistency.

It is important to note that, after the HPE, the number of extra heavy
flavors ($N_f$) and their mass parameter ($\kh$) appear only in a single
parameter $h$, Eq.~(\ref{eq:h}).
Because of this, $\kh$ and $\Nf$ have been replaced by $h$ in the
arguments of $R$ and $R'$, and our purpose turns to finding the critical
value of $h$, $h_c$.
It should be also noted that the second derivatives of
Eqs.~(\ref{eq:r-1}) and (\ref{eq:rdash}) with regard to $P$ are
identical because the difference is proportional to $P$.

One side remark is below.
Thus far, we have restricted the extra heavy quarks to be degenerate.
But the extension to the nondegenerate case is straightforward by
interpreting $h$ as $h=2 \sum_{f=1}^{\Nf} (2 \kh)^{N_t}$.
In the following, we only consider the degenerate case for simplicity.

Choosing $\hat X=\hat P$ significantly simplifies the procedure to find
the critical value of $h$~\cite{Ejiri:2007ga,Ejiri:2012rr} as follows.
We are interested in the shape of the potential at the (pseudo)critical
temperature, which requires $\beta_{\rm ref}$ to be tuned to its
(pseudo)critical value, $\beta_{c}$ (or $\beta_{pc}$).
This tuning can be totally skipped if we look at the curvature, {\it
i.e.} the second derivative of the potential with respect to $P$,
because it is independent of $\beta_{\rm ref}$.
$R$ is independent of $\beta_{\rm ref}$ as stated above.
$V_{\rm light}$ depends on $\beta_{\rm ref}$, but its second derivative
does not as explained below.

The finite temperature transition of two-flavor QCD is always a
crossover for the two-flavor masses adopted in this paper.
Then, at any temperatures, the shape of the PDF (or equivalently
histogram) for $\hat P$ in two-flavor QCD can be well approximated,
around the peak, by Gaussian form,
\begin{eqnarray}
   w(P;\beta,\kl,0,0)\big|_{P\sim \bar P(\beta,\kl)}
\propto
   \exp\left[ - \frac{6\,N_{\rm site}\,
   	              (\,P - \bar P(\beta,\kl)\,)^2}
                     {2\,\chi_P(\beta,\kl)}
       \right]\ ,
\end{eqnarray}
where $\bar P(\beta,\kl)=\langle \hat P \rangle_{\beta,\kl}$ is the
average of generalized plaquette at $\beta$ and $\kl$, and $\chi_P$
is the susceptibility of $P$, given by
\begin{eqnarray}
  \chi_P(\beta,\kl)
= 6 N_{\rm site}
  \langle\, (\,\hat P - \bar P(\beta,\kl)\, )^2\,
  \rangle_{\beta,\kl}\ .
\end{eqnarray}
Substituting this into Eq.~(\ref{eq:V-two-flavor}) yields, up to a
constant shift,
\begin{eqnarray}
      V_{\rm light}(P; \beta_{\rm ref},\kl)
      \big|_{P\sim \bar P(\beta,\kl)}
&=&   \frac{6\,N_{\rm site}\,
          (\, P - \bar P(\beta,\kl)\,)^2}
	 {2\,\chi_P(\beta,\kl)}
    - 6(\beta_{\rm ref} - \beta)\,N_{\rm site}\,P
    \ .
\end{eqnarray}
Then, the first and second derivatives are given by
\begin{eqnarray}
      \frac{d V_{\rm light}(P;\beta_{\rm ref},\kl)}{dP}
      \bigg|_{P\sim \bar P(\beta,\kl)}
&=&   \frac{6\,N_{\rm site}\,
            (\, P - \bar P(\beta,\kl)\,)}
	   {\chi_P(\beta,\kl)}
    - 6(\beta_{\rm ref} - \beta)\,N_{\rm site}\ ,
\label{eq:d1V0}
\\
      \frac{d^2 V_{\rm light}(P;\kl)}{dP^2}
      \bigg|_{P\sim \bar P(\beta,\kl)}
&=&   \frac{6\,N_{\rm site}}{\chi_P(\beta,\kl)}\ .
\label{eq:d2V0}
\end{eqnarray}
Thus, we can calculate the curvature of the two-flavor part
by collecting $\chi_P(\beta,\kl)$ obtained at various $\beta$.
Importantly, Eq.~(\ref{eq:d2V0}) is independent of
$\beta_{\rm ref}$.
In summary, the curvature of the total effective potential
\begin{eqnarray}
  \frac{d^2 V(P;\beta_{\rm ref},\kl,h)}{dP^2}
= \frac{d^2 V_{\rm light}(P;\beta_{\rm ref},\kl)}{dP^2}
- \frac{d^2 \ln R(P; \kl, h)}{dP^2}\ ,
\label{eq:curvature}
\end{eqnarray}
is independent of $\beta_{\rm ref}$.

The procedure to identify $h_c$ in the chiral limit of two flavors goes
as follows.
At $h=0$, the contribution of the extra heavy flavors is trivially zero,
and the system is reduced to two-flavor QCD, where the transition is a
crossover.
Therefore, the second derivative of the potential is always positive.
As $h$ is increased from zero, the minimum of the curvature takes zero
at some point, which gives $h_c$.
In this procedure, one needs not tune $\beta_{\rm ref}$ to
$\beta_{\rm pc}$ or $\beta_{\rm c}$, because the curvature is
independent of the temperature or $\beta_{\rm ref}$.
This simplification does not occur in general, and one such example
is explicitly shown in Sec.~\ref{subsec:polyakov}.
By looking at the light quark mass dependence of $h_c(\kl)$, we try to
extract $h_c$ in the chiral limit.

\section{Numerical results}
\label{sec:numerical_results}
\subsection{Simulation parameters}
\label{subsec:lat-para}

\begin{figure}[tb]
\begin{center}
\begin{tabular}{cc}
\includegraphics*[width=0.5 \textwidth,clip=true]
{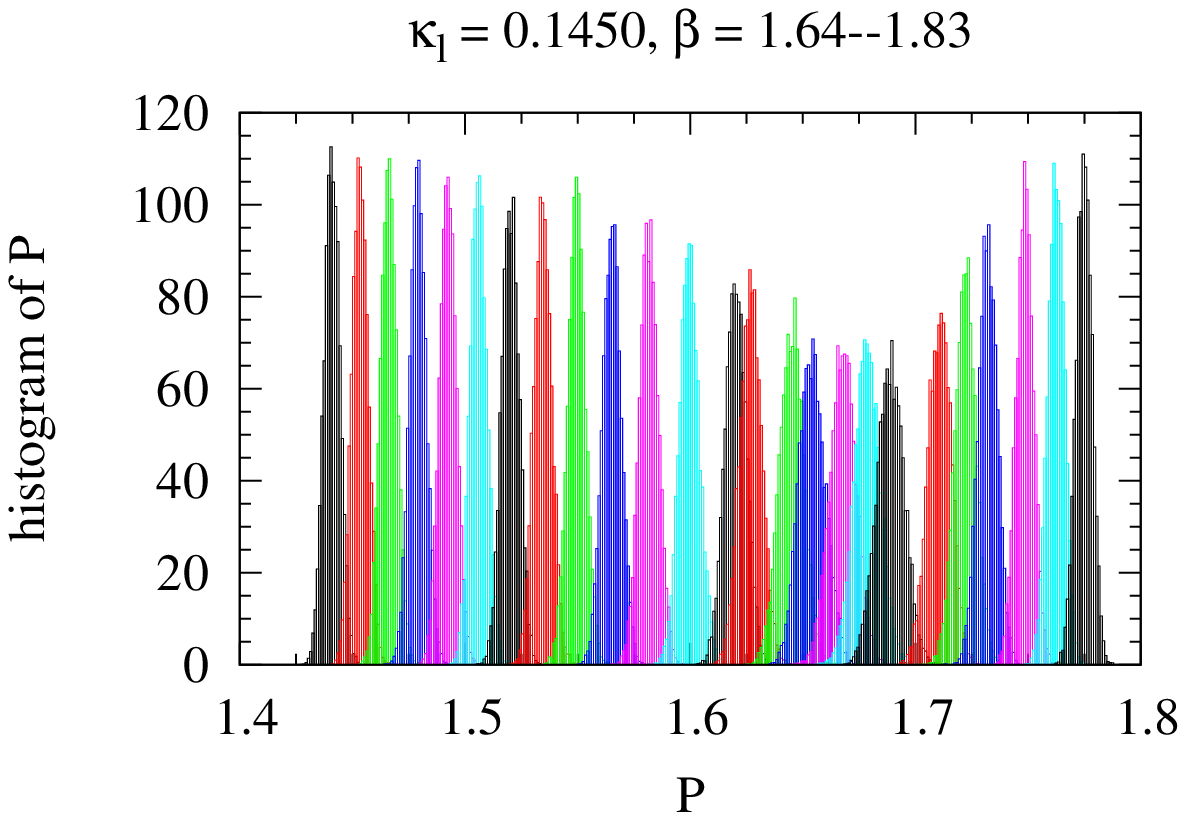}& 
\includegraphics*[width=0.5 \textwidth,clip=true]
{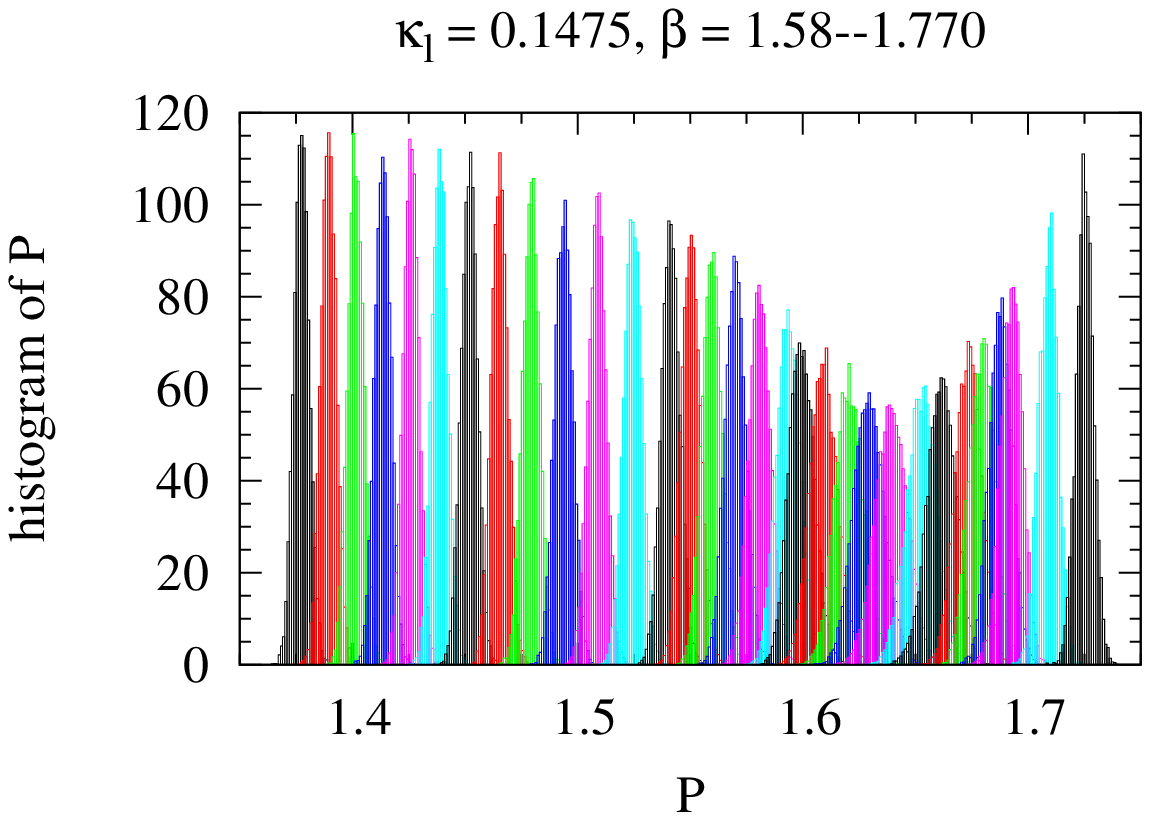}\\
\includegraphics*[width=0.5 \textwidth,clip=true]
{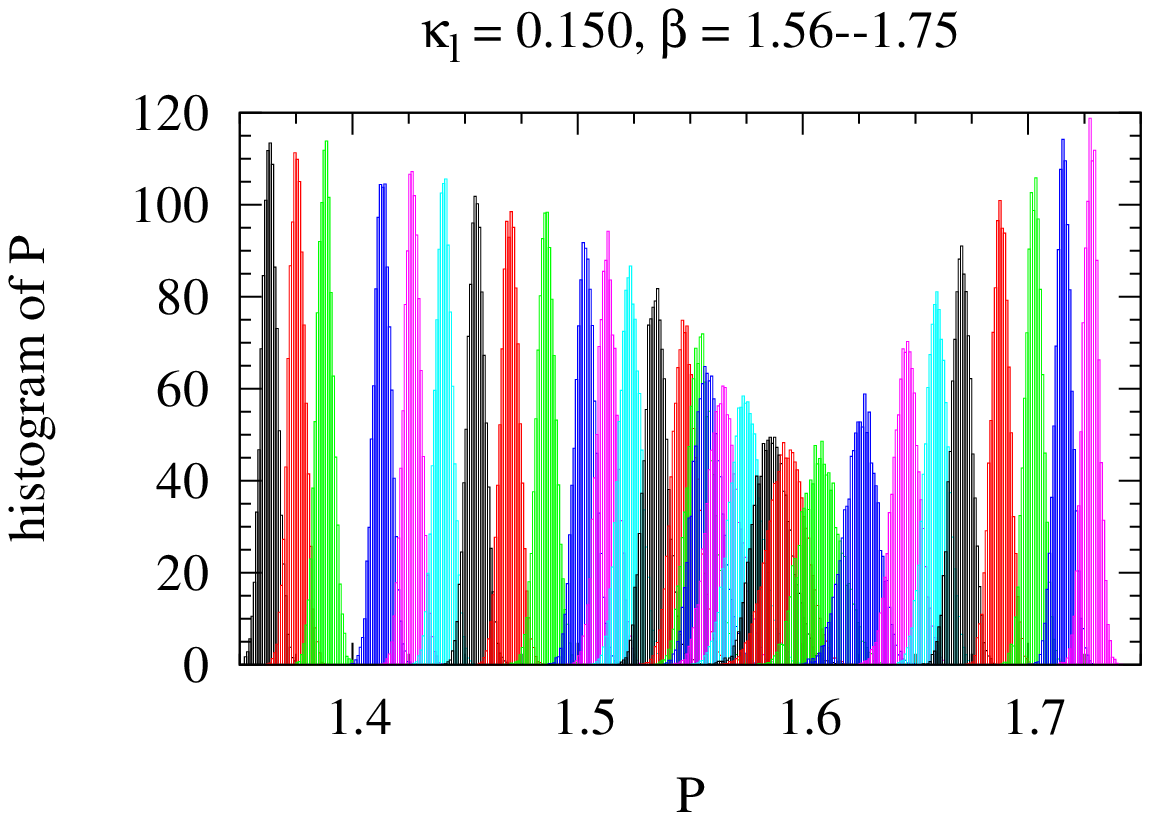}&
\includegraphics*[width=0.5 \textwidth,clip=true]
{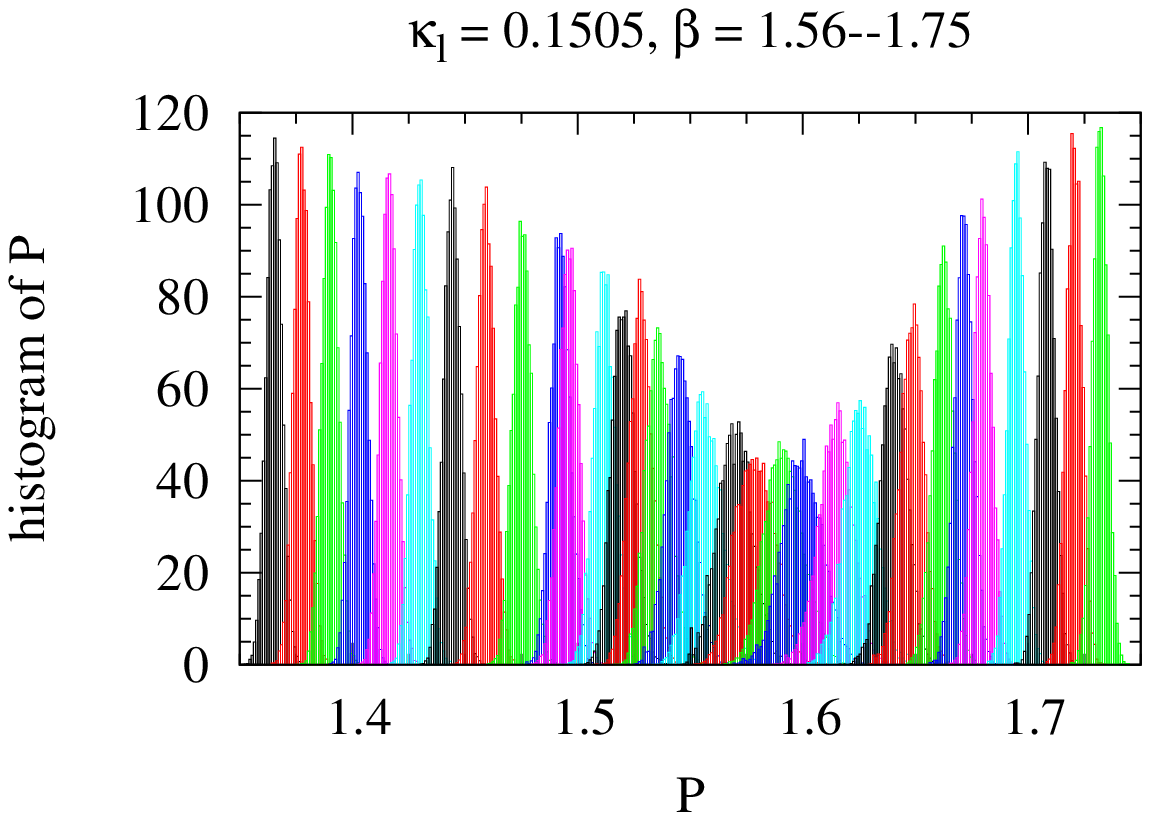}\\
\end{tabular}
\vspace{-1ex}
\caption{Histogram of the generalized plaquette at four values of
 $\kl$.}
\label{fig:histogram-action}
\end{center}
\end{figure}

Following Ref.~\cite{whot10}, we take the Iwasaki gauge action
($c_1=-0.331$) and the $O(a)$-improved Wilson fermion action with the
perturbatively improved $c_{\rm sw}$ for two flavors of light quarks.
Simulations are performed on $N_{\rm site}=16^3\times 4$ lattices with
25 to 32 $\beta$ values at each of four $\kl$,
and 10,000 to 40,000 trajectories have been accumulated at each
simulation point.
Four light quark masses are ranging from $\kl=0.145$ to 0.1505.
Figure \ref{fig:histogram-action} shows the histogram of the generalized
plaquette.

\begin{table}[t]
 \centering
 \begin{tabular}{cc|ccccc}
 $\kl$ & $c_{\rm sw}$ & $\beta_{\rm pc}$ & $a m_\pi$ & $a m_\rho$
& $m_\pi/m_\rho$ & $a m_{\rm pcac}$\\
  \hline
   0.1450 & 1.650 & 1.778 &0.779(1) & 1.172(\ 2)& 0.665(33)
          & 0.0535(1)\\
   0.1475 & 1.677 & 1.737 & 0.651(1) & 1.130(\ 5)& 0.576(28)
          & 0.0350(2)\\
   0.1500 & 1.707 & 1.691 & 0.514(2) & 1.099(10) & 0.468(24)
          & 0.0202(2)\\
   0.1505 & 1.712 & 1.681 & 0.495(2) & 1.082(13) & 0.458(23)
          & 0.0186(2)\\
 \end{tabular}
 \caption{Simulation parameters ($\kl$ and $c_{\rm sw}$) and the
 pseudocritical $\beta$ ($\beta_{\rm pc}$) in two-flavor QCD,
 determined from the peak of the susceptibility for the generalized
 plaquette.
 $m_\pi/m_\rho$ and other quantities are determined on a $16^3\times 32$
 lattice at $\beta_{\rm pc}$ for each $\kl$.}
 \label{tab:spectrum}
\end{table}

At the pseudocritical point $\beta_{\rm pc}$ for each $\kl$, we carried
out zero temperature simulations on $16^3\times 32$ lattices to find the
mass ratio of pseudoscalar and vector mesons, $m_\pi/m_\rho$, and the
quark mass defined through the partially conserved axialvector current
(PCAC),
\begin{eqnarray}
    am_{\rm pcac}
&=& \frac{\langle \sum_{\vec x}
          (A_4(N_t/2+1,\vec x)-A_4(N_t/2-1,\vec x)) P(0)\rangle}
         {4\, \langle \sum_{\vec x}P(N_t/2,\vec x) P(0) \rangle}\ ,
\end{eqnarray}
where $A_4(x)$ and $P(x)$ are the flavor nonsinglet, local axialvector
and pseudoscalar operator, respectively, and $N_t=32$.
Then, the four values of $\kl$ in this paper turn out to cover
$0.46 < m_\pi/m_\rho < 0.66$, or $0.019 < am_{\rm pcac} < 0.054$.
These results are tabulated in Table~\ref{tab:spectrum}.

\subsection{Main results}
\label{subsec:main-result}

We present the numerical results for the two terms in the right-hand
side of Eq.~(\ref{eq:curvature}) separately, and focus on the second
term first.
With the approximated $\delta$ function (\ref{eq:delta-func}),
$\ln R(P;\kl,h)$ [Eq.~(\ref{eq:r-1})] and $\ln R'(P;\kl,h)$
[Eq.~(\ref{eq:rdash})] are calculated.
We take two values of $\Delta$, 0.0001, 0.00025, to see the
stability of the results, and the discrepancy arising from a different
choice of $\Delta$ is taken as the systematic uncertainties.
In the following plots, the results with $\Delta=0.0001$ are shown
unless otherwise stated.

\begin{figure}[t]
\begin{center}
\begin{tabular}{cc}
\includegraphics*[width=0.5 \textwidth,clip=true]
{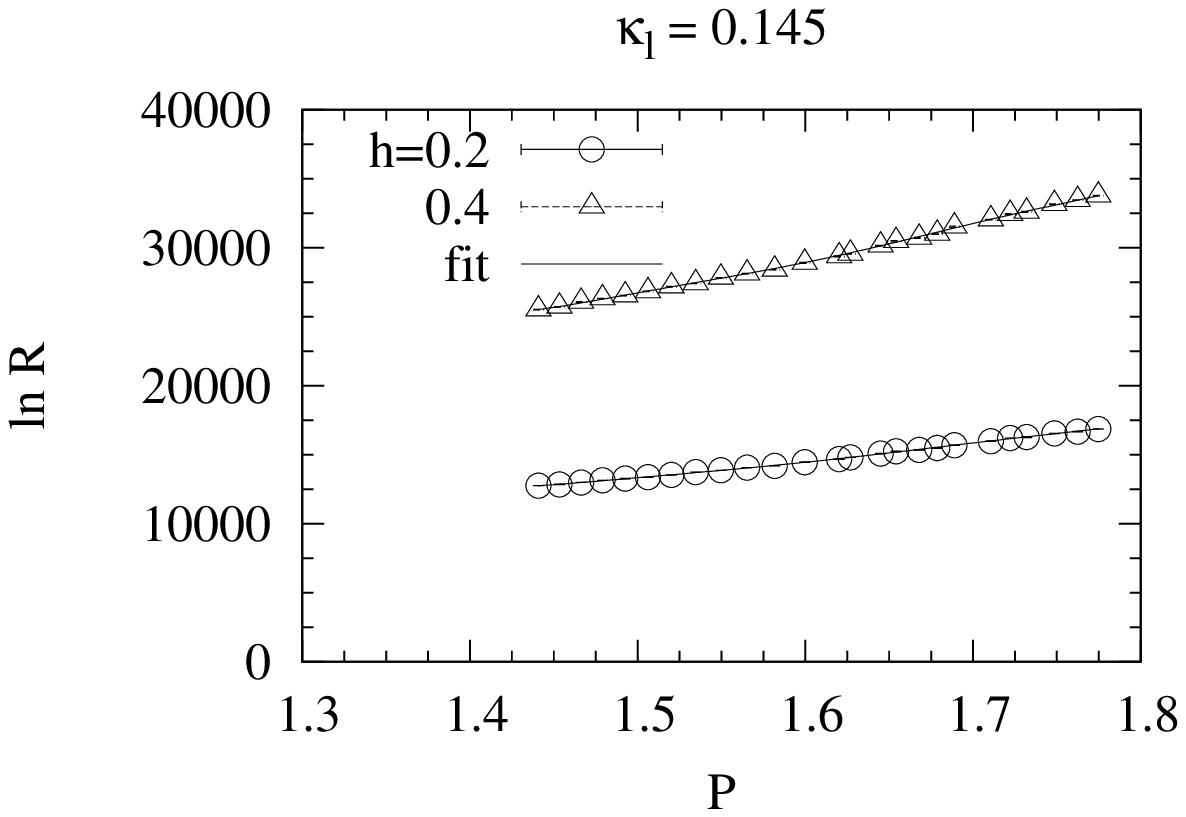}&
\includegraphics*[width=0.5 \textwidth,clip=true]
{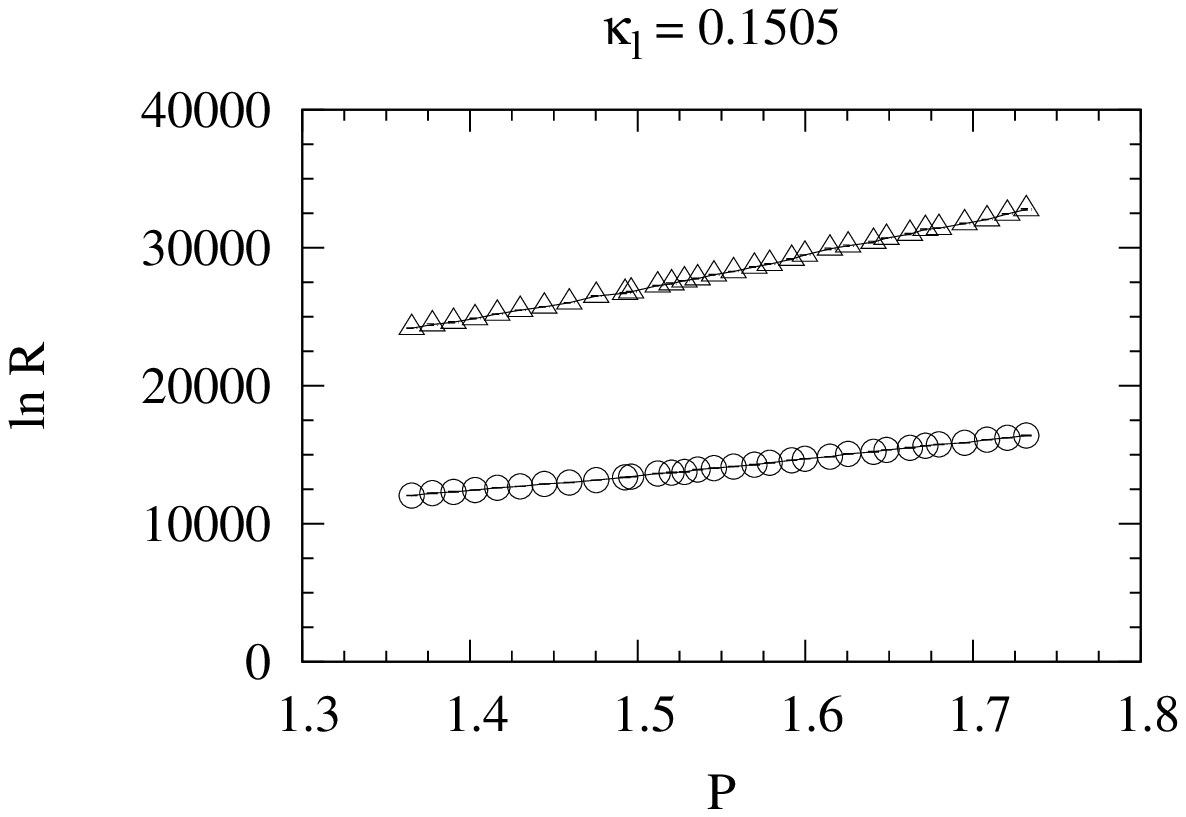}\\[-2ex]
 (a)&(b)\\[2ex]
\includegraphics*[width=0.5 \textwidth,clip=true]
{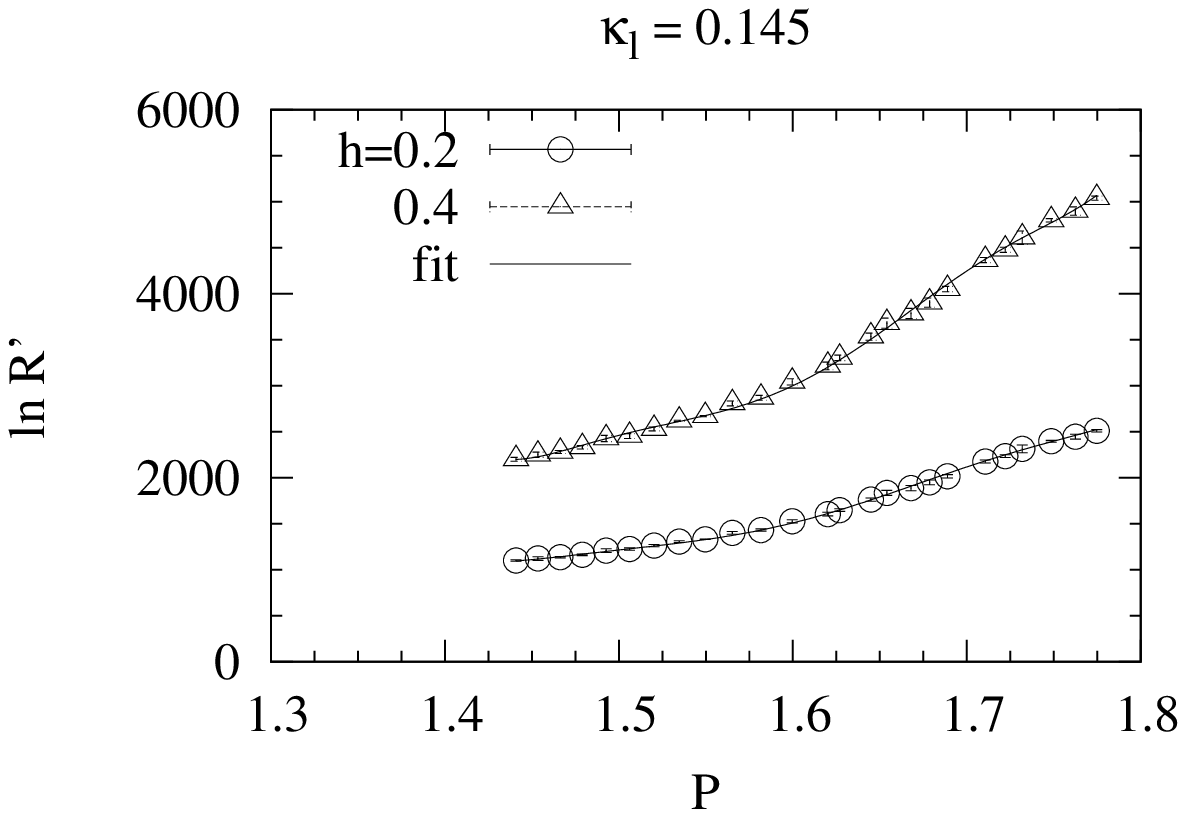}&
\includegraphics*[width=0.5 \textwidth,clip=true]
{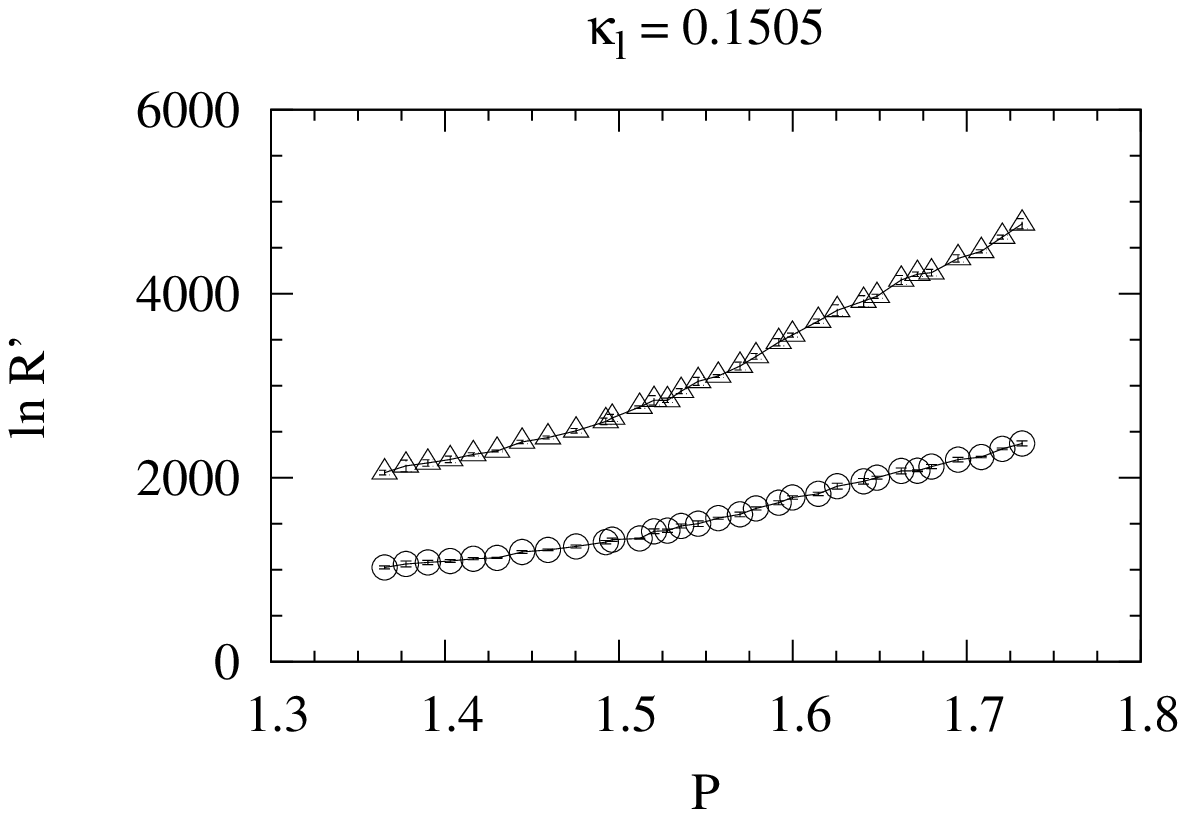}\\[-2ex]
 (c)&(d)\\
\end{tabular}
\vspace{-1ex}
\caption{$P$ dependence of $\ln R$ [Eq.~(\ref{eq:r-1})]
 at $\kl=0.145$ (a) and 0.1505 (b),
 and $\ln R'$ [Eq.~(\ref{eq:rdash})] at $\kl=0.145$ (c) and 0.1505 (d).
 The results at $h$=0.2 to 0.4 are shown.
 }
\label{fig:lnr}
\vspace{-1ex}
\end{center}
\end{figure}

The $P$ dependence of $\ln R(P;\kl,h)$ and $\ln R'(P;\kl,h)$ are shown
in Fig.~\ref{fig:lnr}.
The statistical errors are invisible on this scale.
The sizes of $\ln R$ and $\ln R'$ differ by an order of magnitude, which
indicates that the difference proportional to $P$ is large and explains
why the curvature in $\ln R$ is less clear than that in $\ln R'$.
The curvature in $\ln R$ and $\ln R'$ originates from the rapid increase
of the Polyakov loop ($\hat L$) contained in Eqs.~(\ref{eq:Y}) or
(\ref{eq:Z}) around $\beta_{\rm pc}$.

We then calculate the slope and curvature by fitting the data
of $\ln R$ and $\ln R'$.
Here let us mention the fits and the selection of the results.
Each data point shown in Fig.~\ref{fig:lnr} is obtained in a separate
simulation, and is totally independent of other points.
The data points are fit to polynomial functions of $P$ as
\begin{eqnarray}
    \ln {R^{(')}}^{\rm fit}(P)
&=& \sum_{i=0}^{N_{\rm poly}} c_i\, P^i \ ,
\label{eq:fit-func}
\end{eqnarray}
over three fit ranges (or three different numbers of the data points),
two different polynomial orders, and two values of $\Delta$.
Since not all the fits are successful, we only keep the fit results,
which satisfy $\chi^2/{\rm dof} < 3$, in the following analysis.

\begin{figure}[t]
\begin{center}
\vspace*{-3ex}
\begin{tabular}{cc}
\includegraphics*[width=0.5 \textwidth,clip=true]
{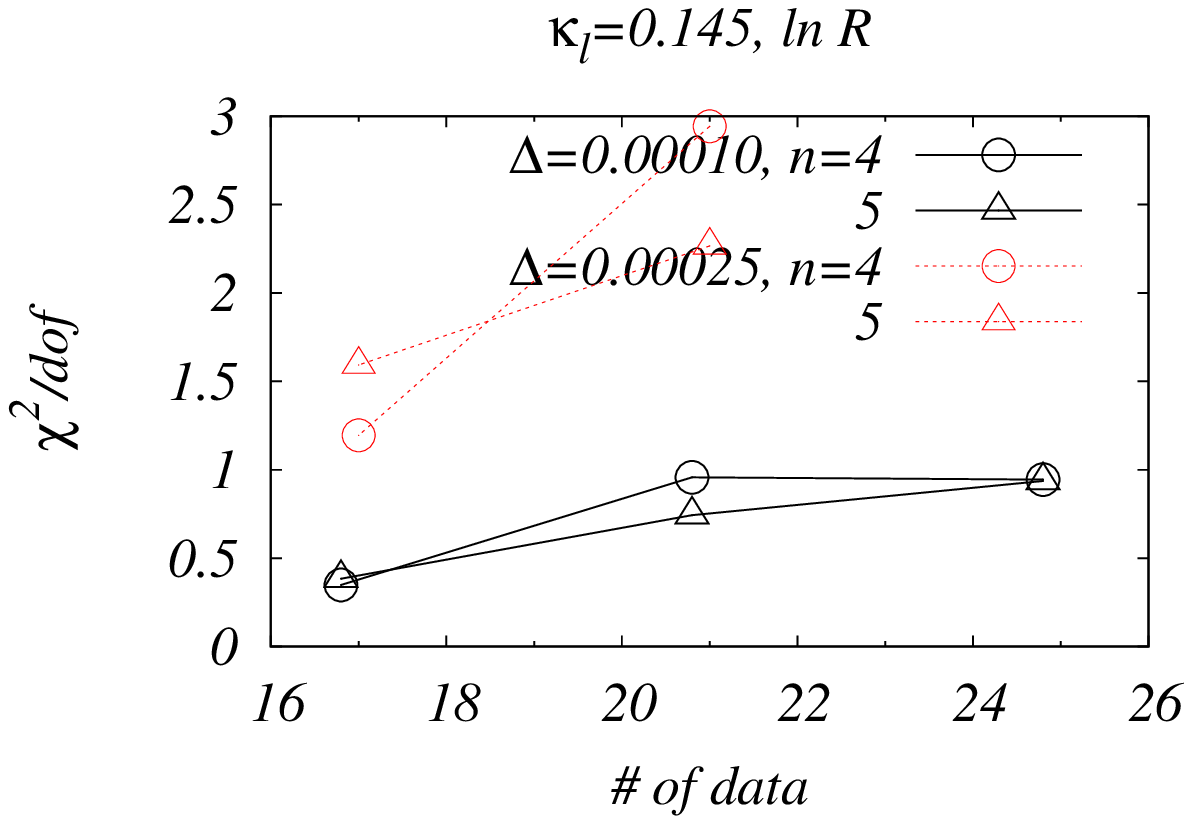}&
\includegraphics*[width=0.5 \textwidth,clip=true]
{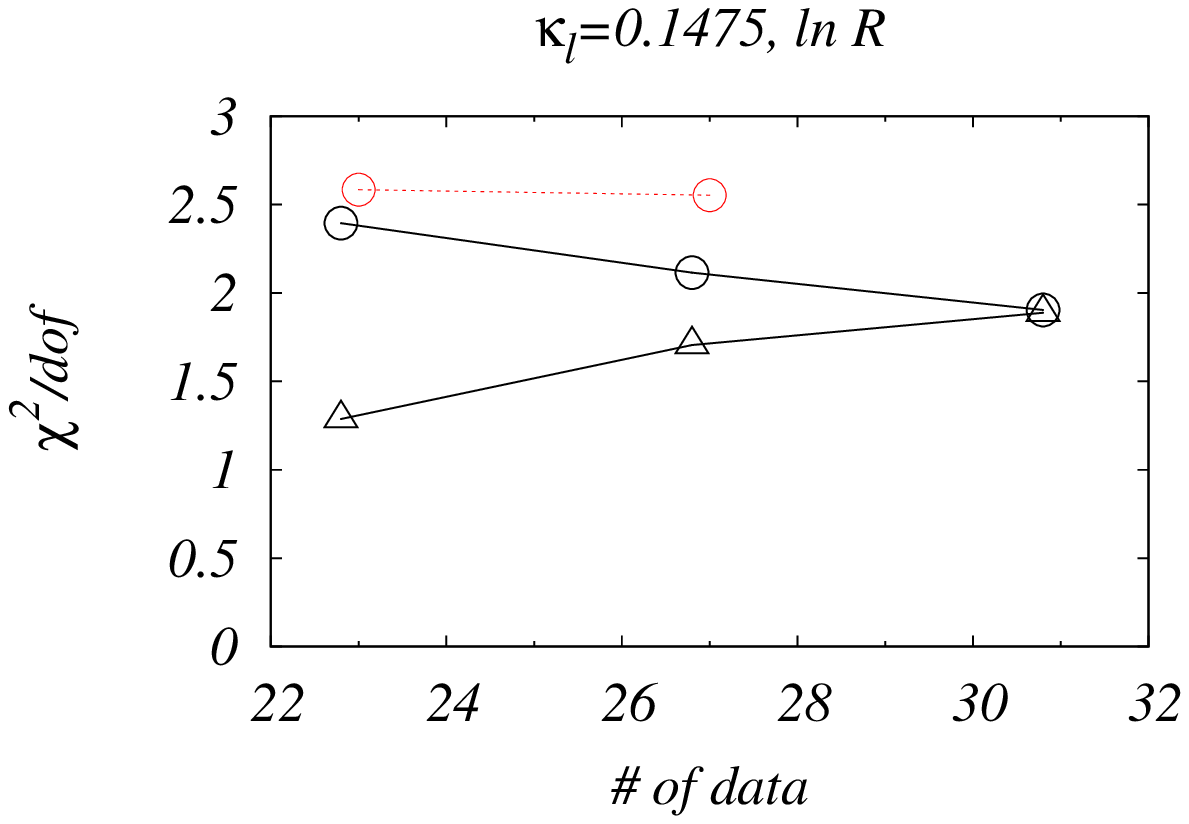}\\[-2ex]
\includegraphics*[width=0.5 \textwidth,clip=true]
{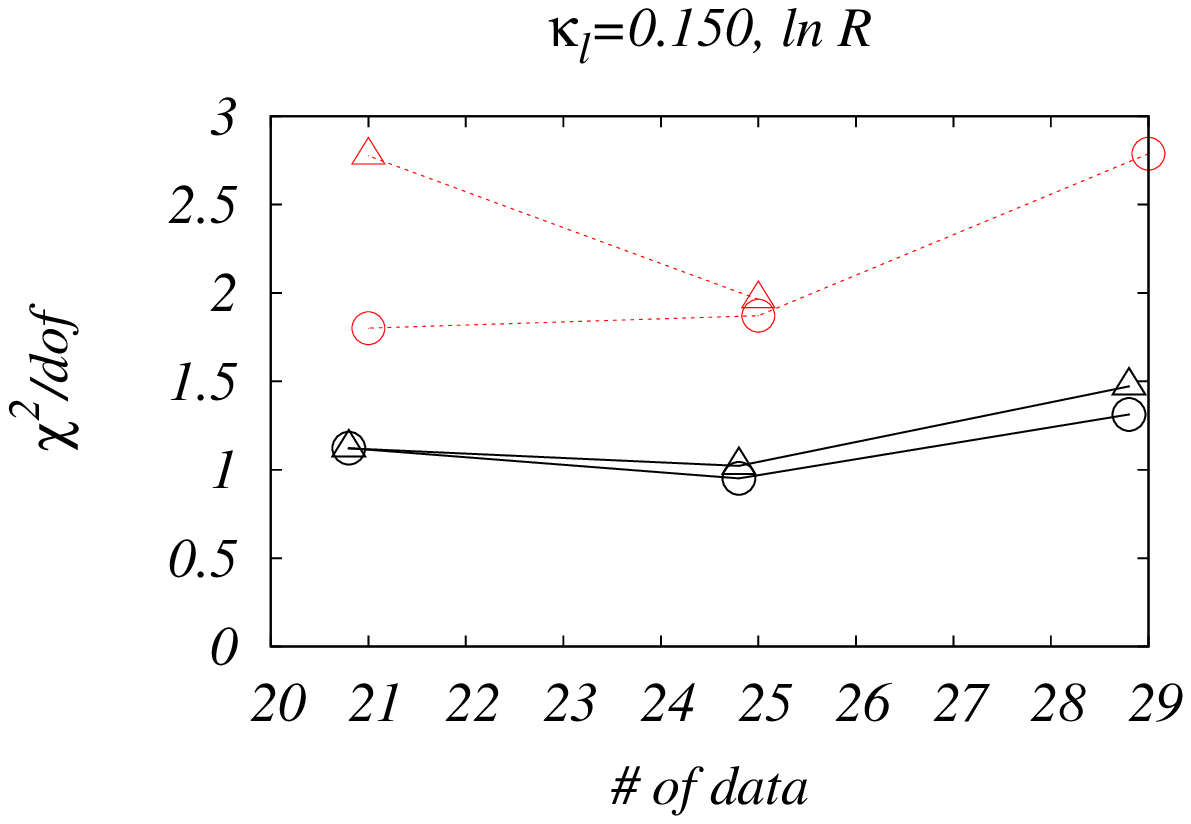}&
\includegraphics*[width=0.5 \textwidth,clip=true]
{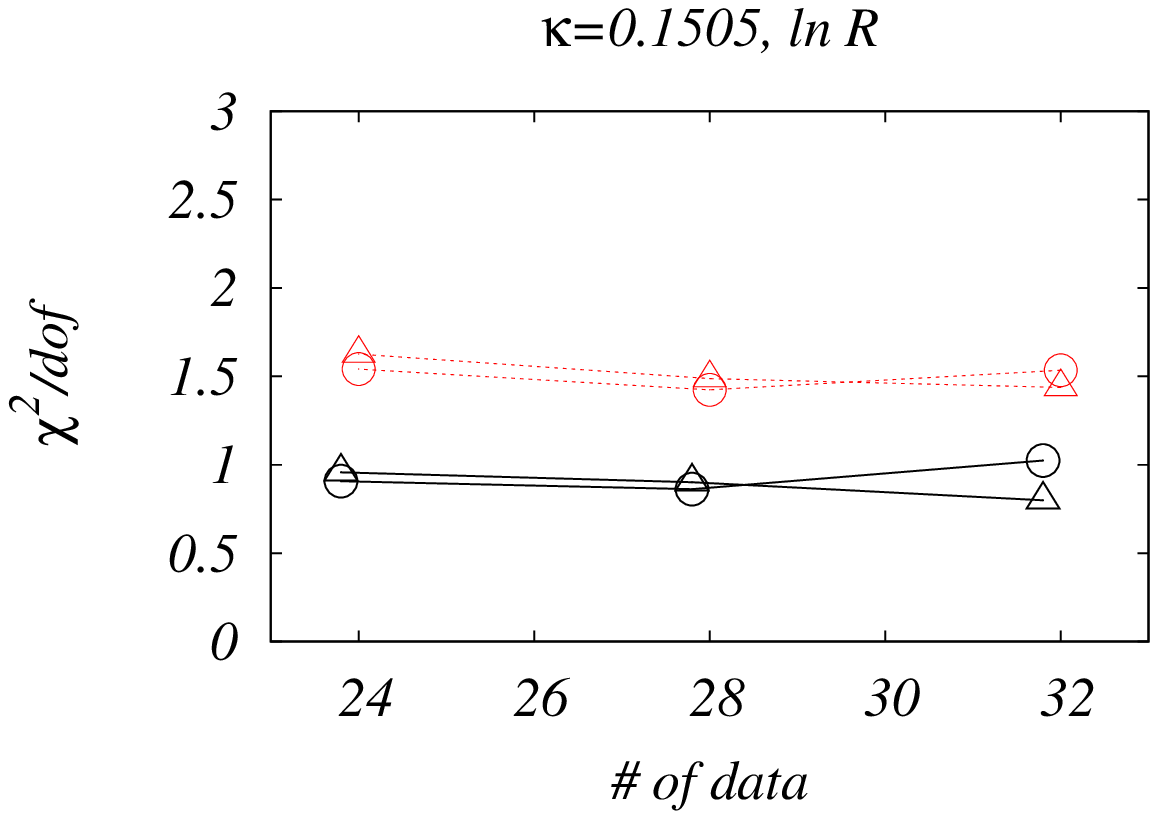}\\[-2ex]
\includegraphics*[width=0.5 \textwidth,clip=true]
{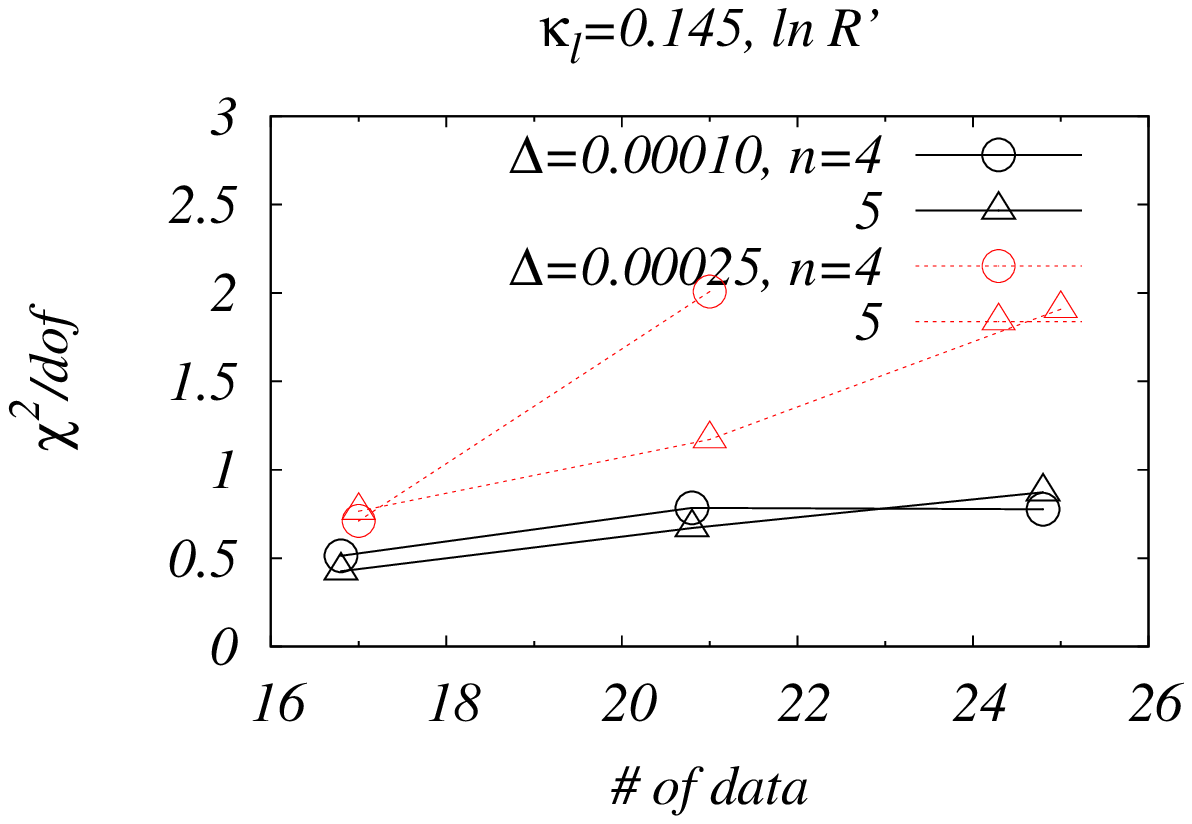}&
\includegraphics*[width=0.5 \textwidth,clip=true]
{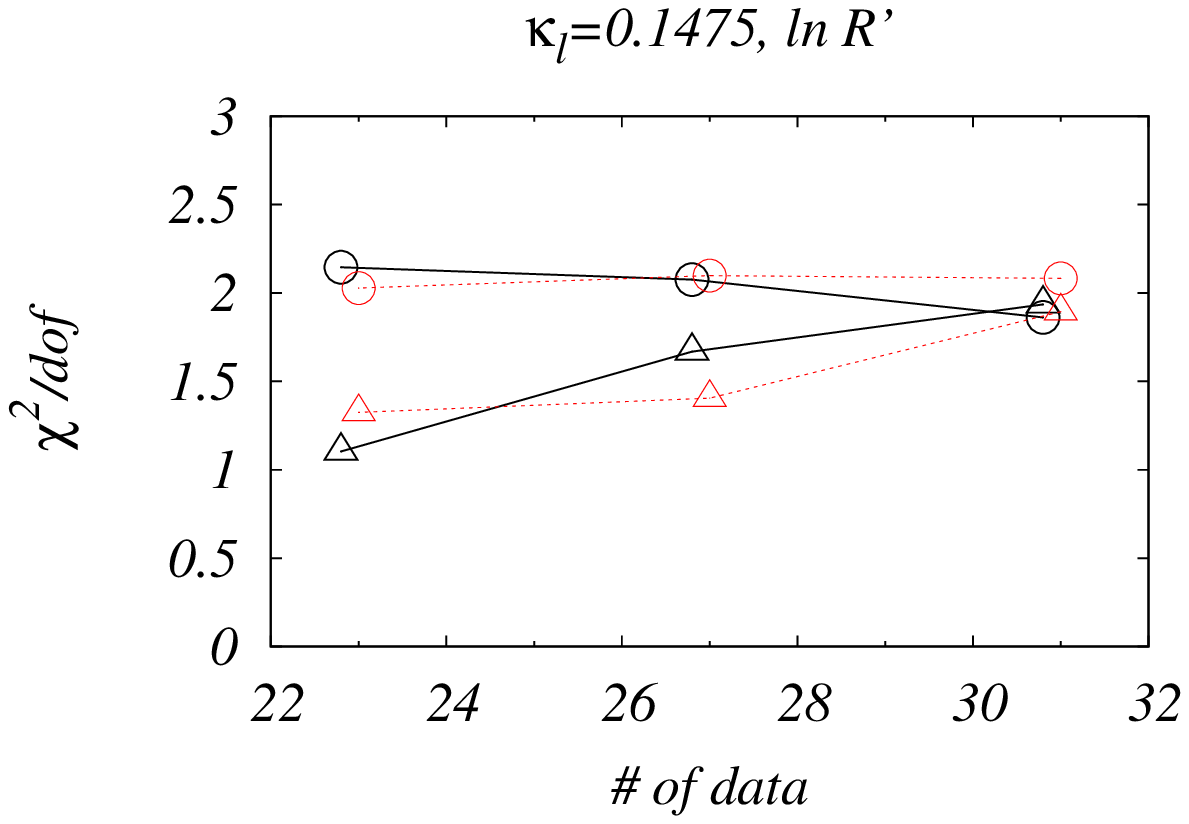}\\[-2ex]
\includegraphics*[width=0.5 \textwidth,clip=true]
{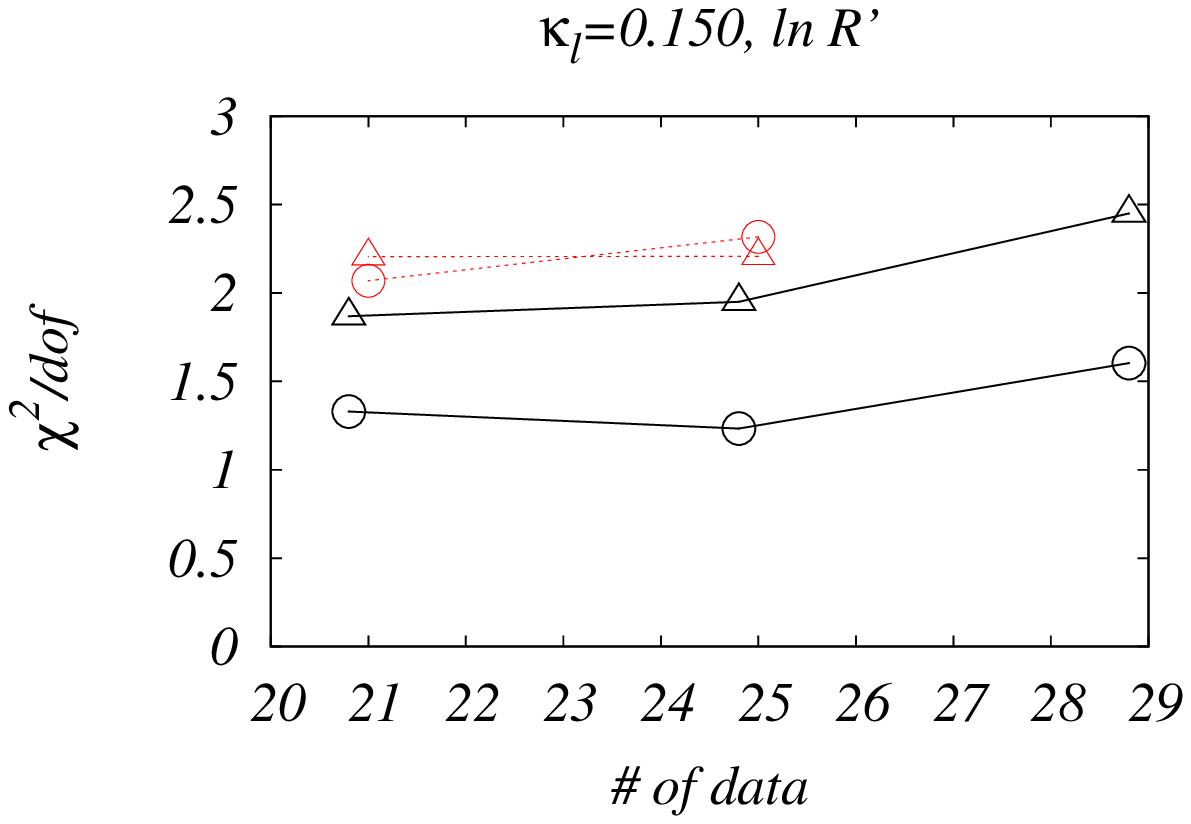}&
\includegraphics*[width=0.5 \textwidth,clip=true]
{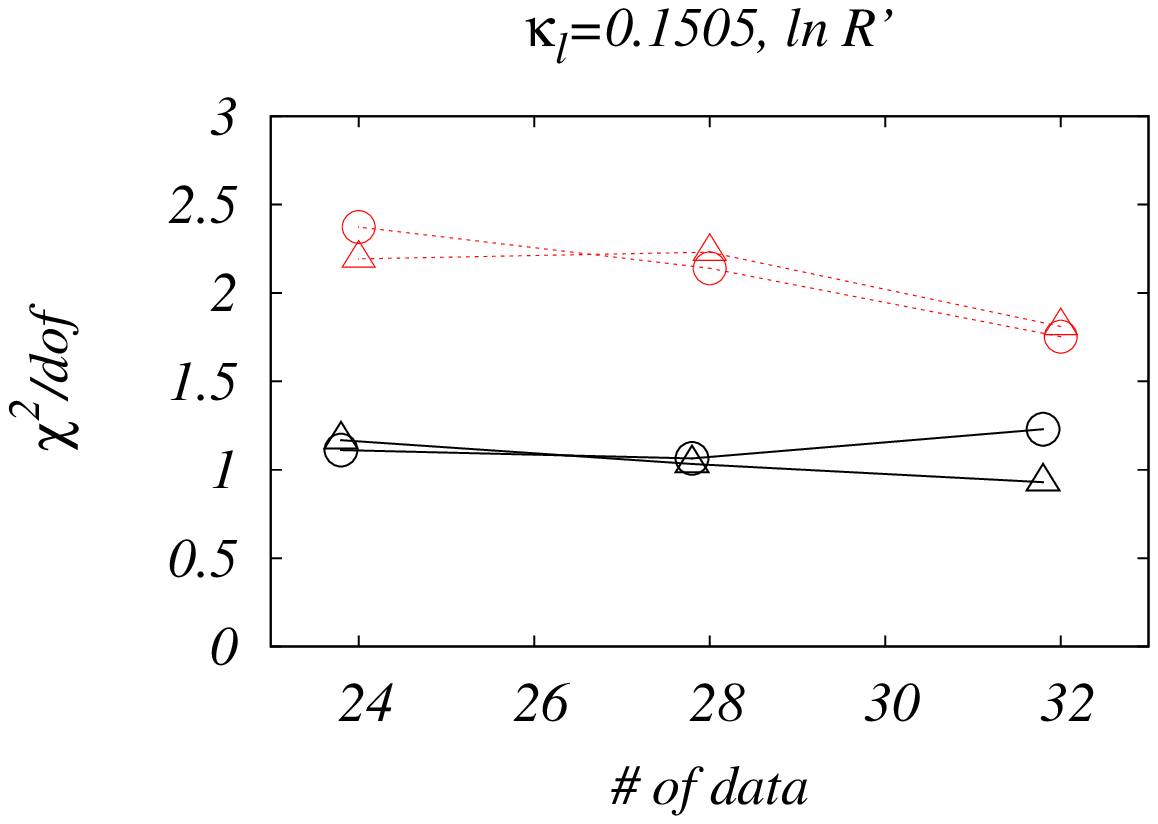}\\[-2ex]
\end{tabular}
\vspace{-3ex}
\caption{The resulting $\chi^2$/dof as a function of the number of data
 points.
 The results for the $n$-th order polynomial fit are shown.}
\label{fig:chi2}
\end{center}
\end{figure}

Figure~\ref{fig:chi2} shows the values of $\chi^2/{\rm dof}$ obtained
from the fit of $\ln R$ and $\ln R'$ as a function of the number of the
data points used, where only the results with $\chi^2/{\rm dof}<3$ are
plotted.
It is seen that $\chi^2/{\rm dof}$ with $\Delta=0.00025$ is always
larger than that with $\Delta=0.0001$.
Once the fit parameters are determined, it is straightforward to
calculate the curvature of the potential.

\begin{figure}[t]
\vspace*{-1ex}
\begin{center}
\begin{tabular}{cc}
\includegraphics*[width=0.45 \textwidth,clip=true]
{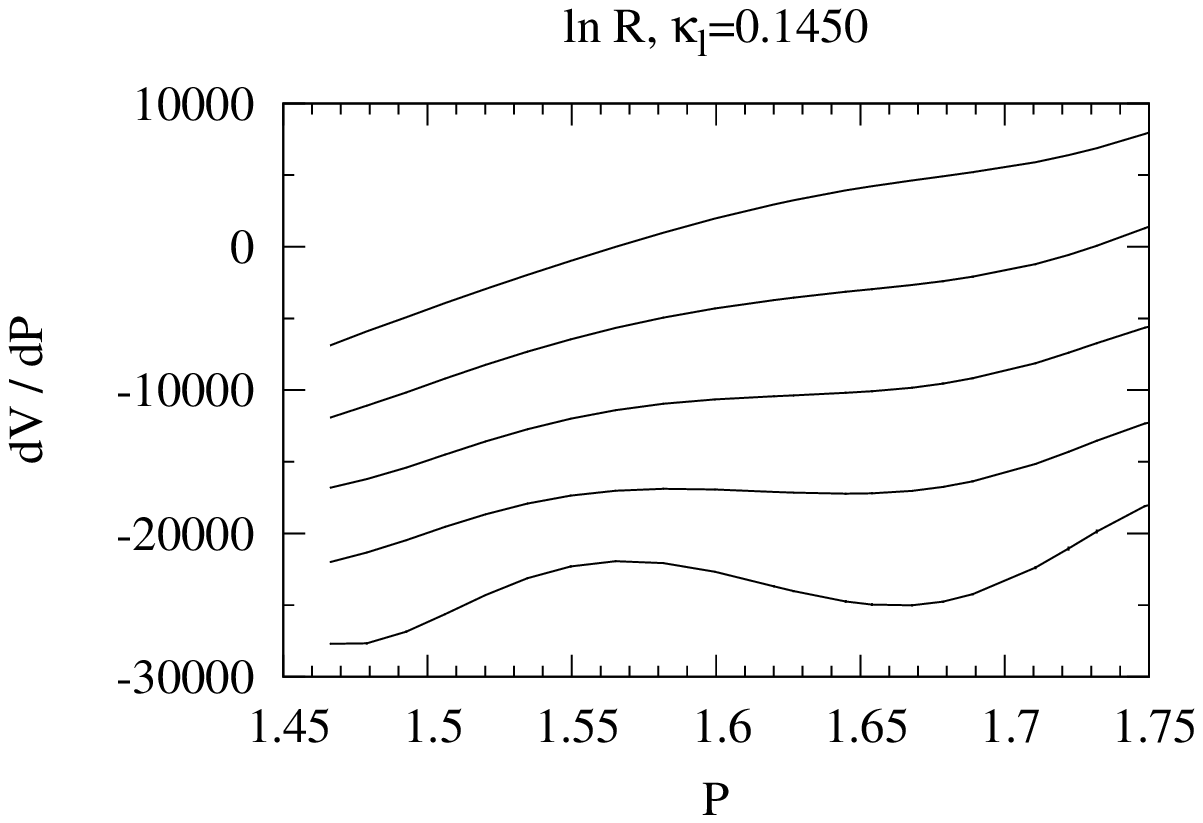}&
\includegraphics*[width=0.45 \textwidth,clip=true]
{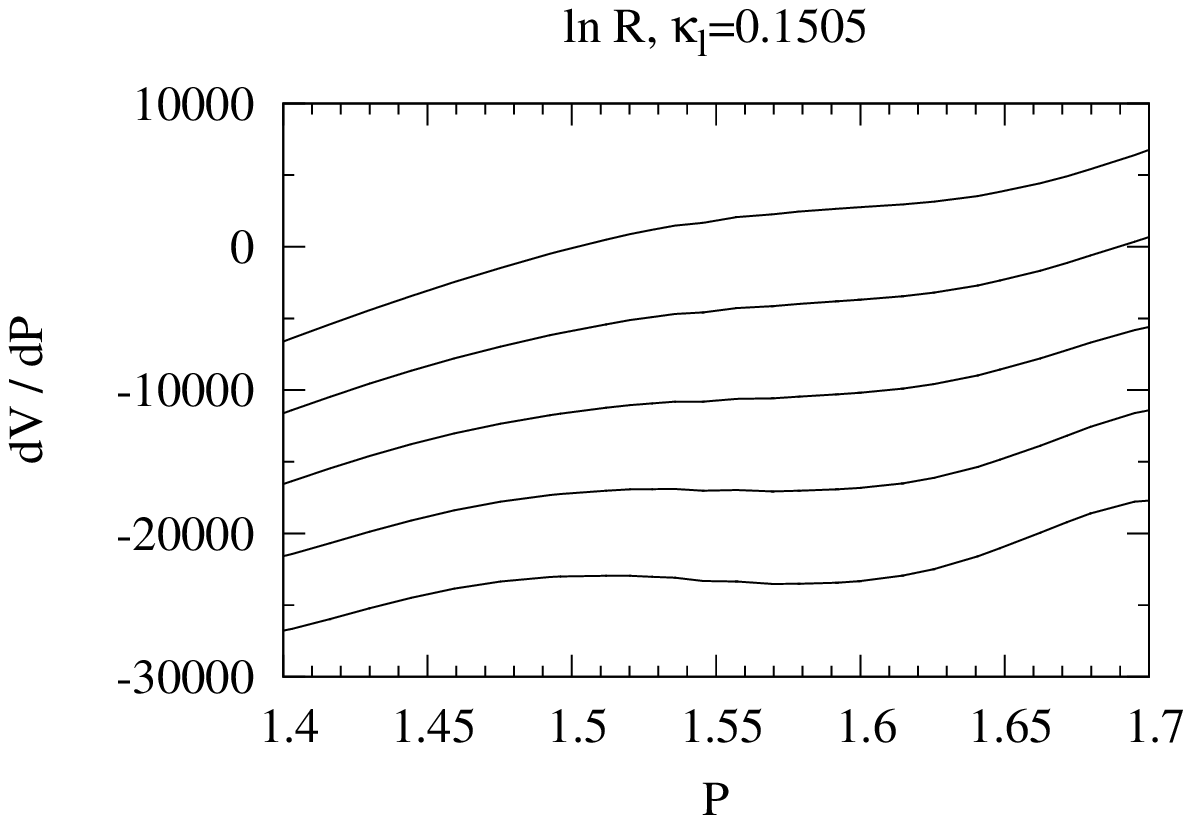}\\
\includegraphics*[width=0.45 \textwidth,clip=true]
{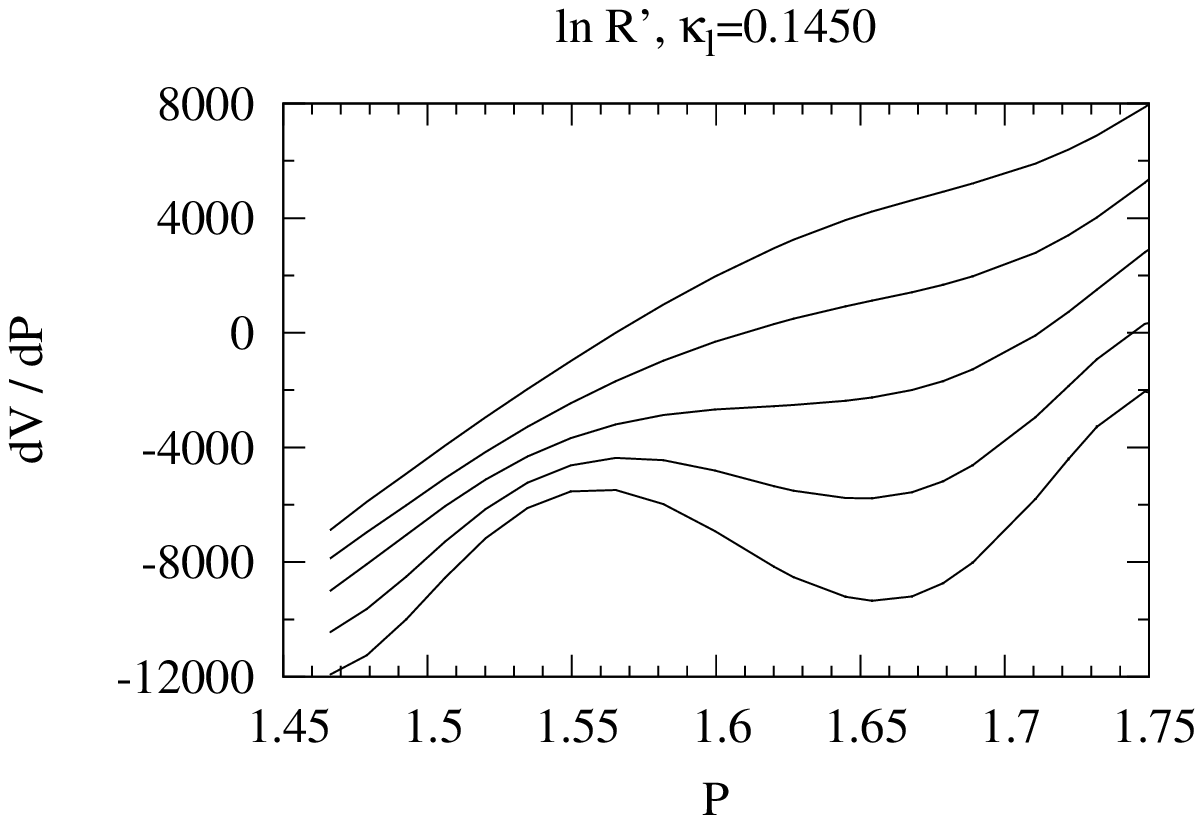}&
\includegraphics*[width=0.45 \textwidth,clip=true]
{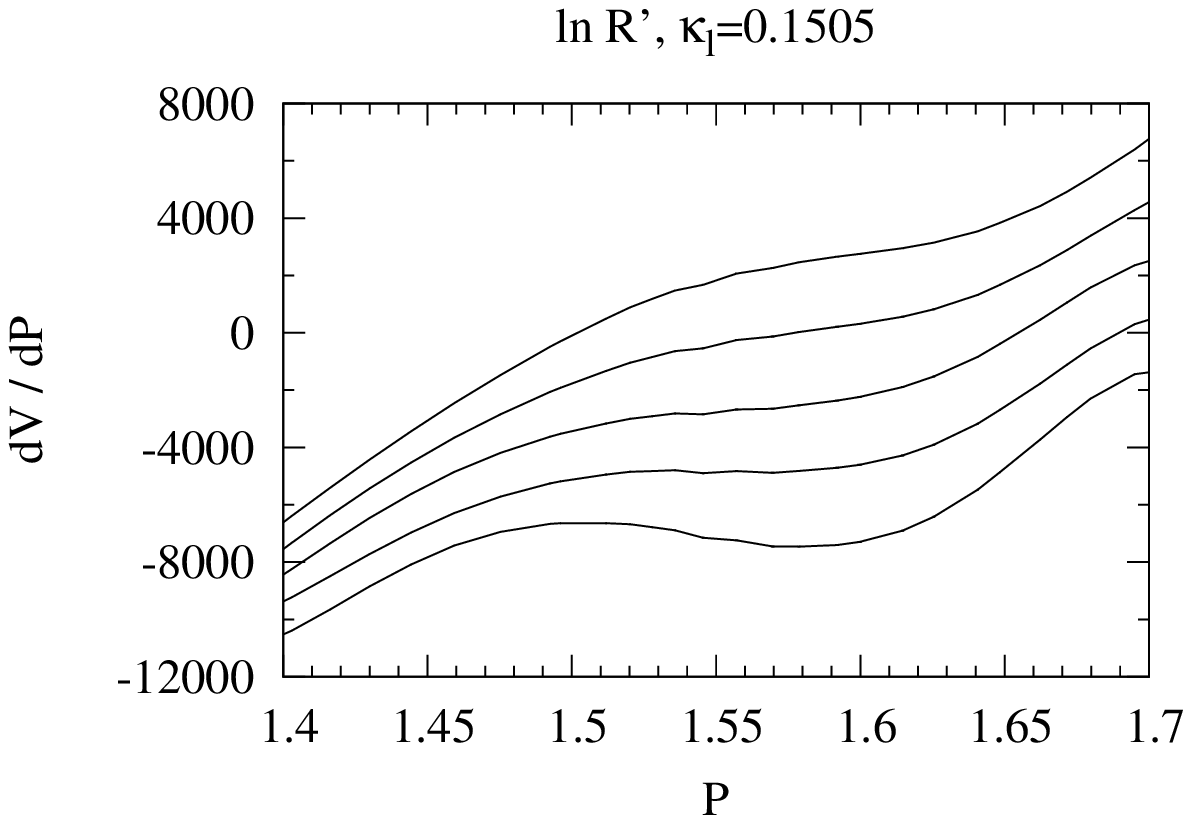}\\
\end{tabular}
\vspace{-1ex}
\caption{The first derivative of the full effective potential is shown
 as a function of $P$.
 $h$ = 0.0, 0.1, 0.2, 0.3, 0.4 from top to bottom.}
\label{fig:dvdp}
 \vspace{-1ex}
\end{center}
\end{figure}

Although we analyze the second derivative to derive the main result,
we discuss the first derivative because it is instructive.
The first derivative of Eq.~(\ref{eq:vefftrans}) is obtained as follows.
First, the numerical values of $\bar P(\beta,\kl)$ and
$\chi_P(\beta,\kl)$ are calculated and substituted in
Eq.~(\ref{eq:d1V0}) to obtain the two-flavor contribution.
Then, the $\Nf$ flavor's contribution, the first derivative of
$\ln R^{(')}$, is determined using the fit results of
Eq.~(\ref{eq:fit-func}).
By adding up, we obtain the first derivative of the full effective
potential.
Figure~\ref{fig:dvdp} shows the typical behavior of the first derivative
of the potential, where the five curves in each plot represent the
results for $h=0.0$, 0.1, $\dots$, 0.4 from top to bottom and the fit
results with $\Delta=0.00010$ and $n=5$ are used.
It is clear that for $h=0$ the curve is monotonically increasing for all
$\kl$ while the ``S'' shape is seen for $h=0.4$.
In principle, it may be possible to determine the critical value of $h$
using these plots, however it is not easy to clearly distinguish an
``S'' shape from a monotonic increase.
Thus, we use the second derivative to determine $h_c$ as described
below.

\begin{figure}[tb]
\begin{center}
\begin{tabular}{cc}
\includegraphics*[width=0.45 \textwidth,clip=true]
{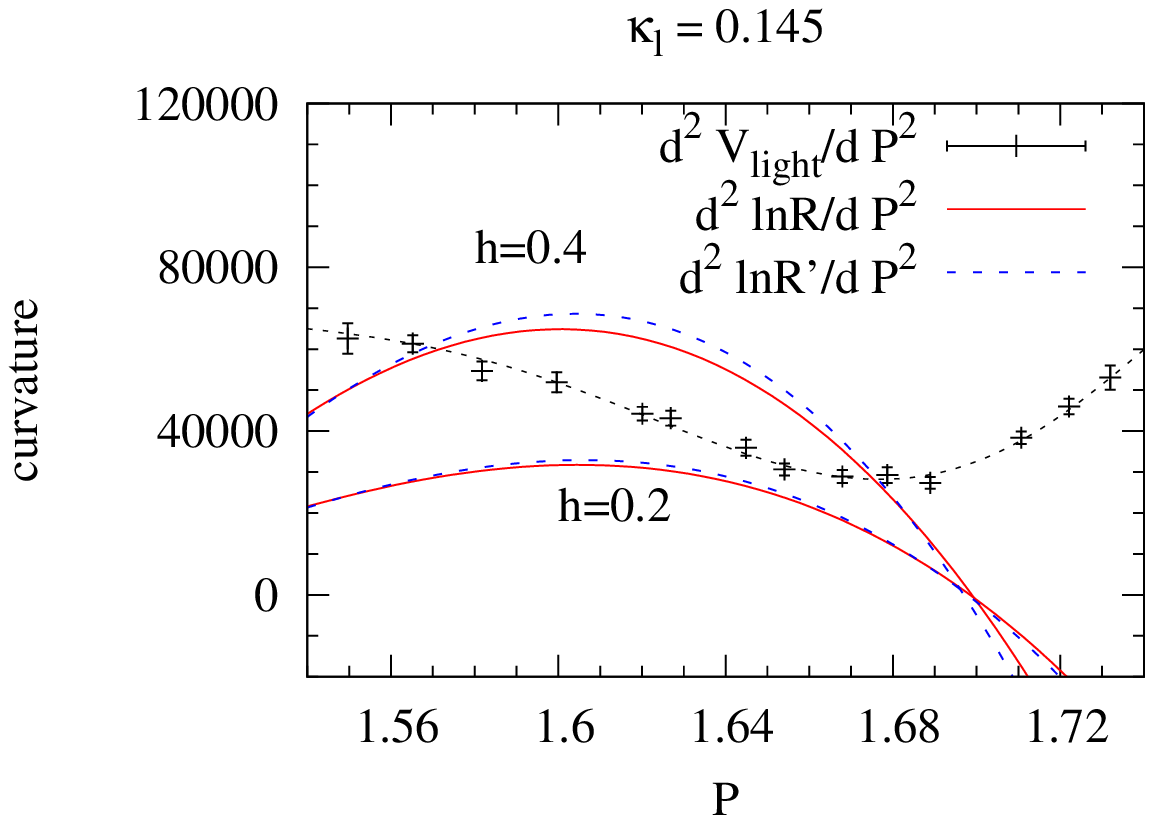}&
\includegraphics*[width=0.45 \textwidth,clip=true]
{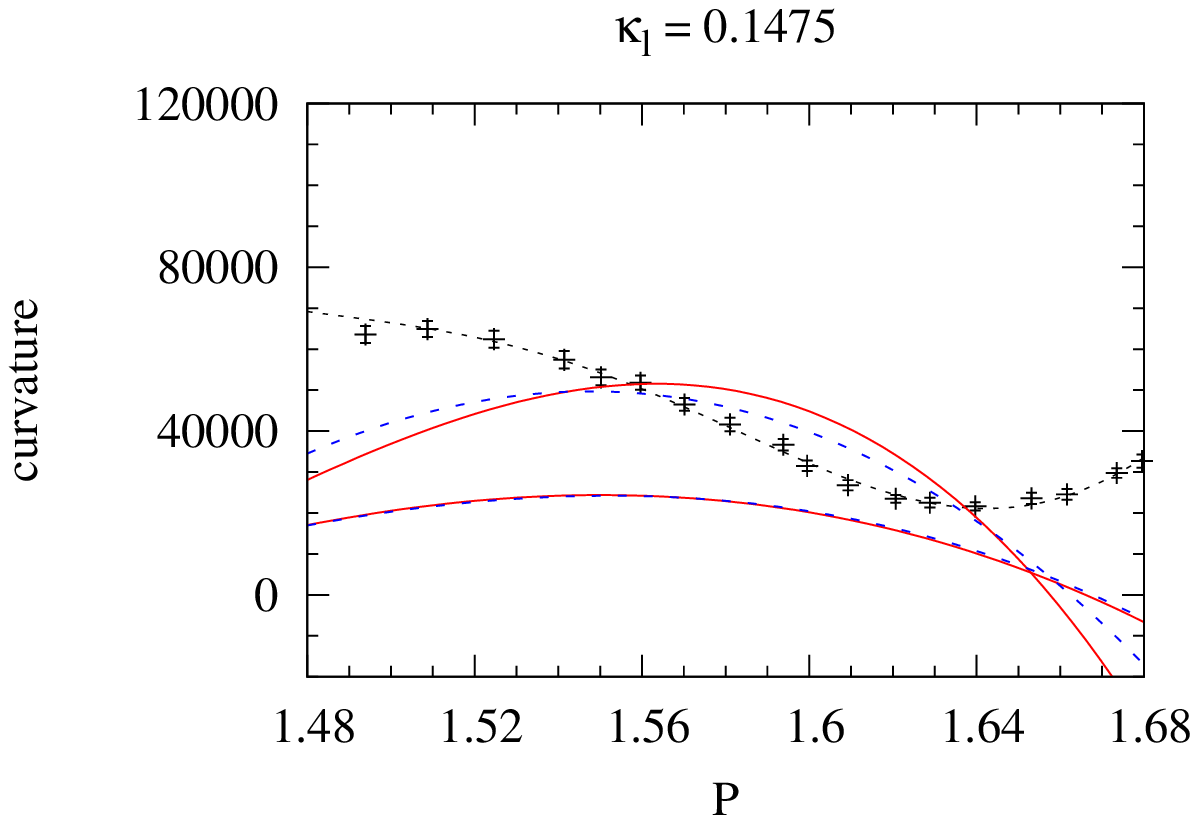}\\
\includegraphics*[width=0.45 \textwidth,clip=true]
{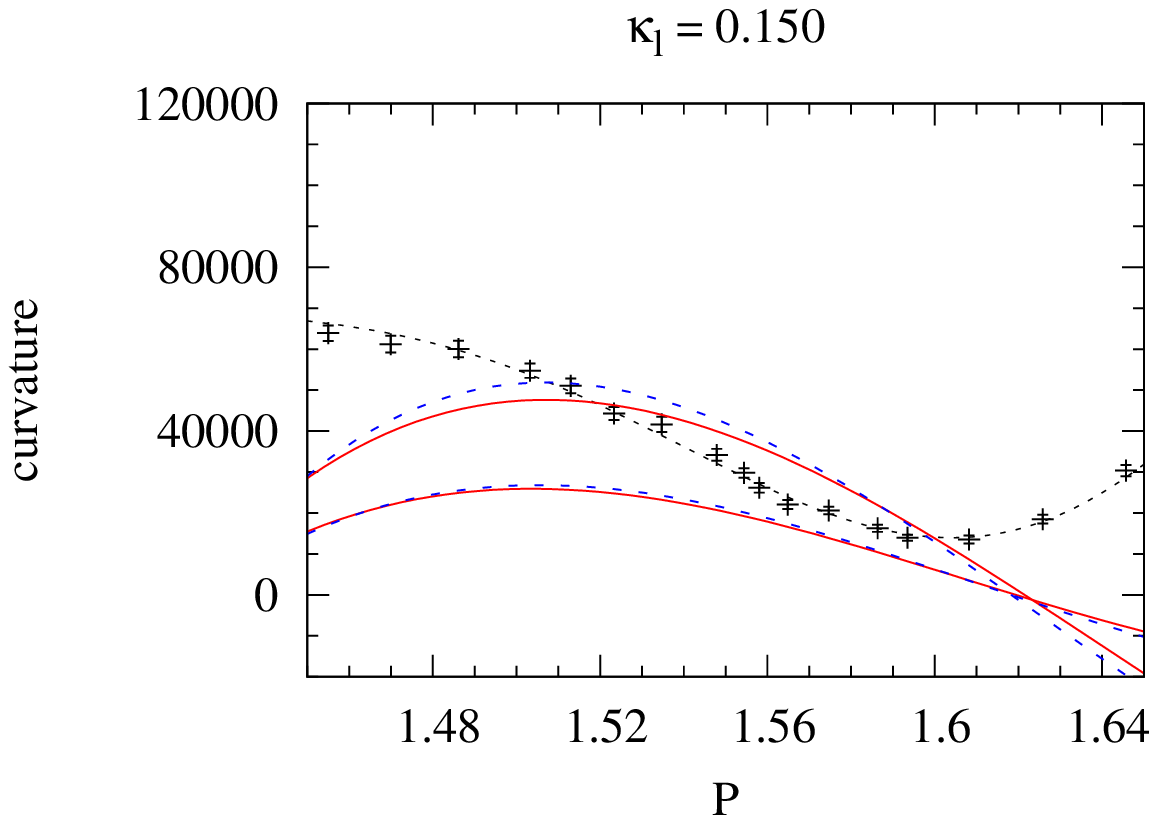}&
\includegraphics*[width=0.45 \textwidth,clip=true]
{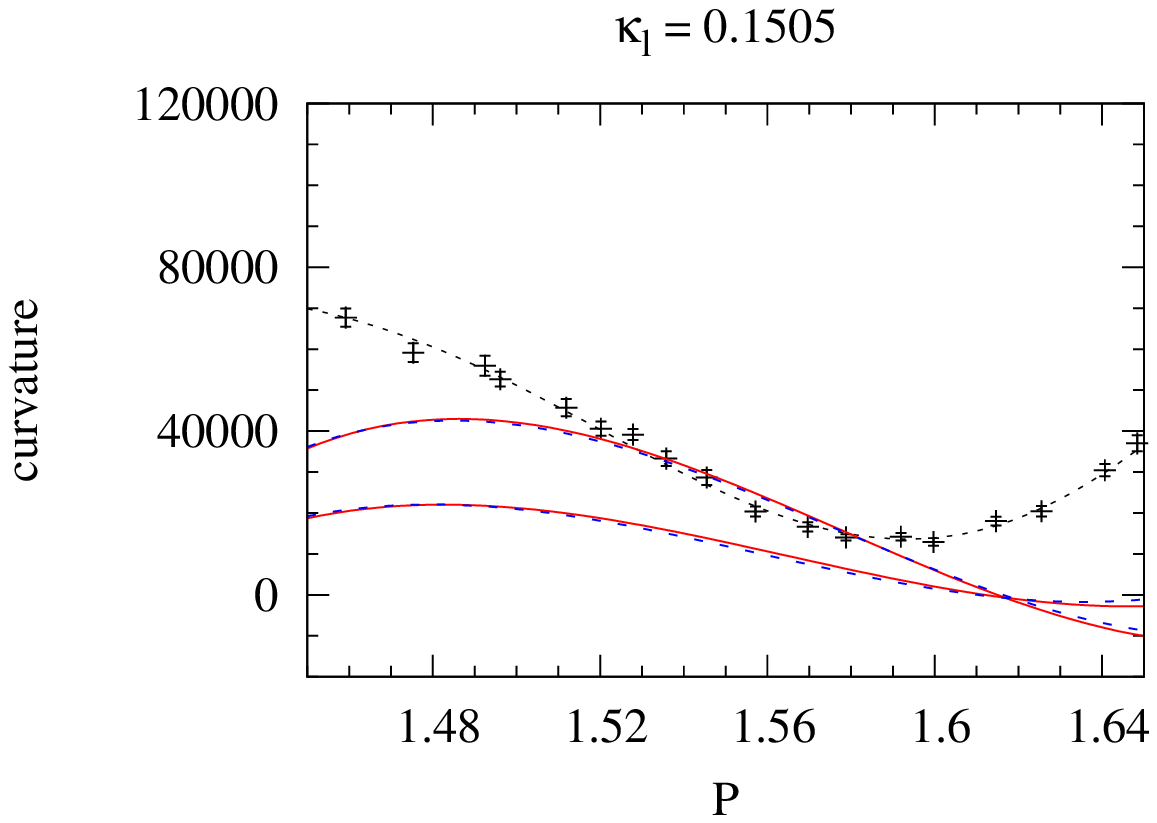}\\
\end{tabular}
\vspace{-2ex}
\caption{The second derivative of the first term and the second term of
 Eq.~(\ref{eq:curvature}) are shown as a function of $P$.
 The second term contribution exceeds that of the first term in a range
 of $P$ when $h=0.4$, which indicates the occurrence of the first order
 transition at such a value of $h$.}
\label{fig:d2lnr}
 \vspace{-2ex}
\end{center}
\end{figure}

The results for the curvature are plotted in Fig.~\ref{fig:d2lnr} for
$\ln R$ (solid curves) and $\ln R'$ (dashed curves), where the results
for $h=0.2$ and 0.4 are shown as examples.
Again, the fit results with $\Delta=0.00010$ and $n=5$ are used.
The difference in the curvature between $\ln R$ and $\ln R'$ turns out
to be reasonably small at all $\kl$.

Next, the curvature of the first term in Eq.~(\ref{eq:curvature}) is
presented, which can be easily calculated using the averaged value and
the susceptibility of $\hat P$ at each $\beta$ as in
Eq.~(\ref{eq:d2V0}).
The curvatures thus obtained are shown in Fig.~\ref{fig:d2lnr} together
with the statistical error, where the fit results obtained with a fifth
order of polynomial are shown by dotted curves.
It is seen that, independently of $\kl$, $d^2V_{\rm light}/dP^2$ is
always positive as expected.

Figure~\ref{fig:d2lnr} shows that $d^2 \ln R/dP^2$ and $d^2 \ln R'/dP^2$
have a peak slightly below the $P$ value at which
$d^2V_{\rm light}/dP^2$ takes the minimum.
This indicates that, in the many flavor system, the phase transition or
rapid crossover occurs at $P$ smaller than the two-flavor case.
For all $\kl$, it is observed that the peak of $d^2 \ln R/dP^2$ or
$d^2 \ln R'/dP^2$ is almost touching the curve of
$d^2V_{\rm light}/dP^2$ at $h=0.2$, and exceeds $d^2V_{\rm light}/dP^2$
at $h=0.4$ in a certain region of $P$.

\begin{figure}[tb]
\begin{center}
\begin{tabular}{cc}
\includegraphics*[width=0.6 \textwidth,clip=true]
{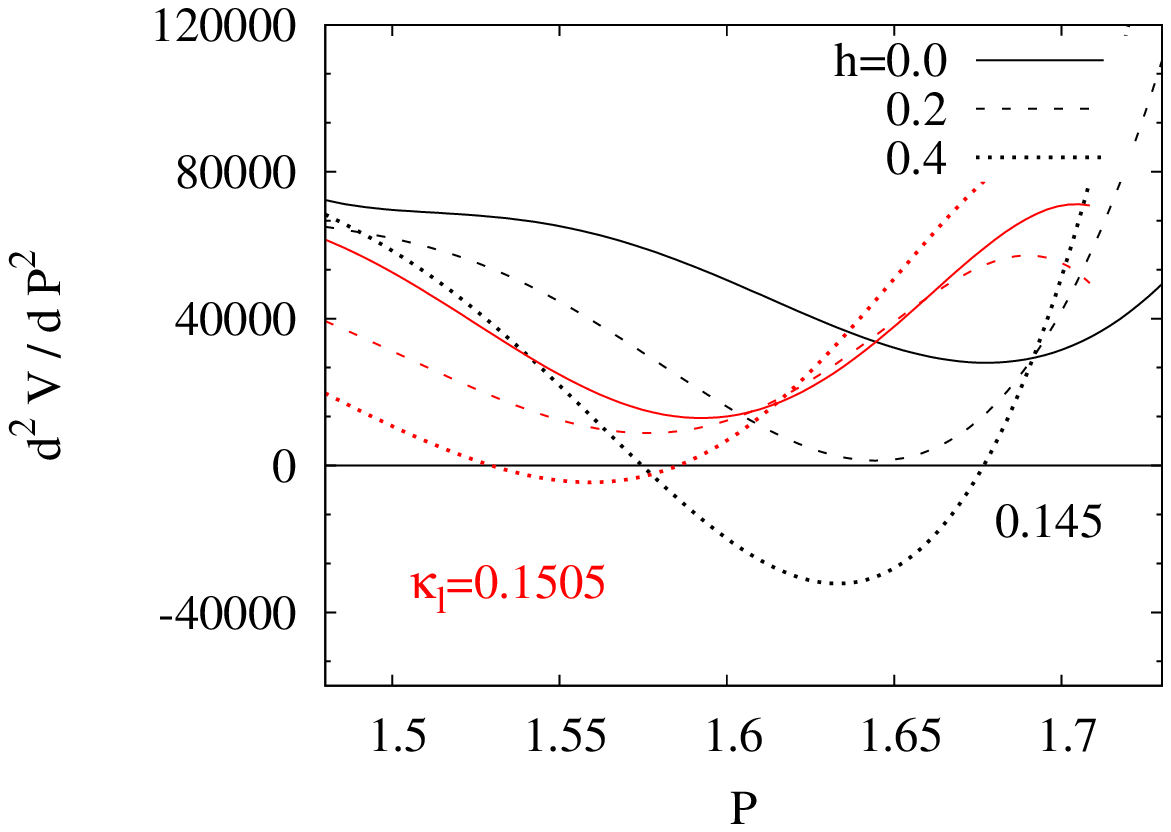}\\
\end{tabular}
\vspace{-2ex}
\caption{The curvature of the effective potential at $\kl=0.1450$ and
 0.1505 and for $h=0.0$ (solid), 0.2 (dashed) and 0.4 (dotted).}
\label{fig:d2Veff-action}
\vspace{-2ex}
\end{center}
\end{figure}

The resulting curvature of the full effective potentials
Eq.~(\ref{eq:curvature}) for $\kl=0.1450$ and 0.1505 are shown in
Fig.~\ref{fig:d2Veff-action}.
It is seen that the minimum of the curvature with $h=0.0$ (solid curve)
approaches to zero towards the chiral limit of two light flavors.
Thus, with this observation alone, one might expect that the potential
with $\kl=0.1505$ requires only a small $h$ to bring it into the double
well shape.
However, at $\kl=0.1505$, it is also true that adding the heavy quarks
does not reduce the minimum of the curvature by much.
As a consequence, $h_c$ takes a similar value at $\kl=0.1450$ and
0.1505.

\begin{figure}[h]
\vspace*{-2ex}
\begin{center}
\begin{tabular}{cc}
\includegraphics*[width=0.5 \textwidth,clip=true]
{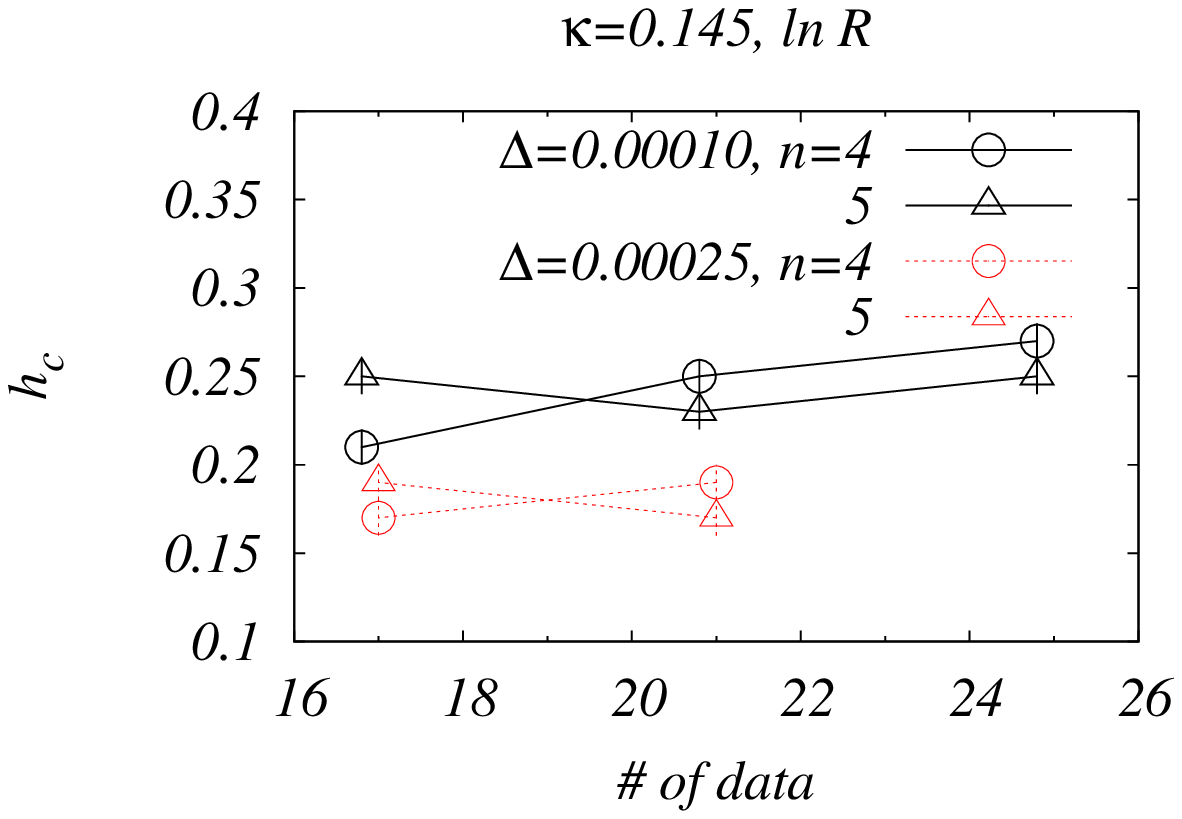}&
\includegraphics*[width=0.5 \textwidth,clip=true]
{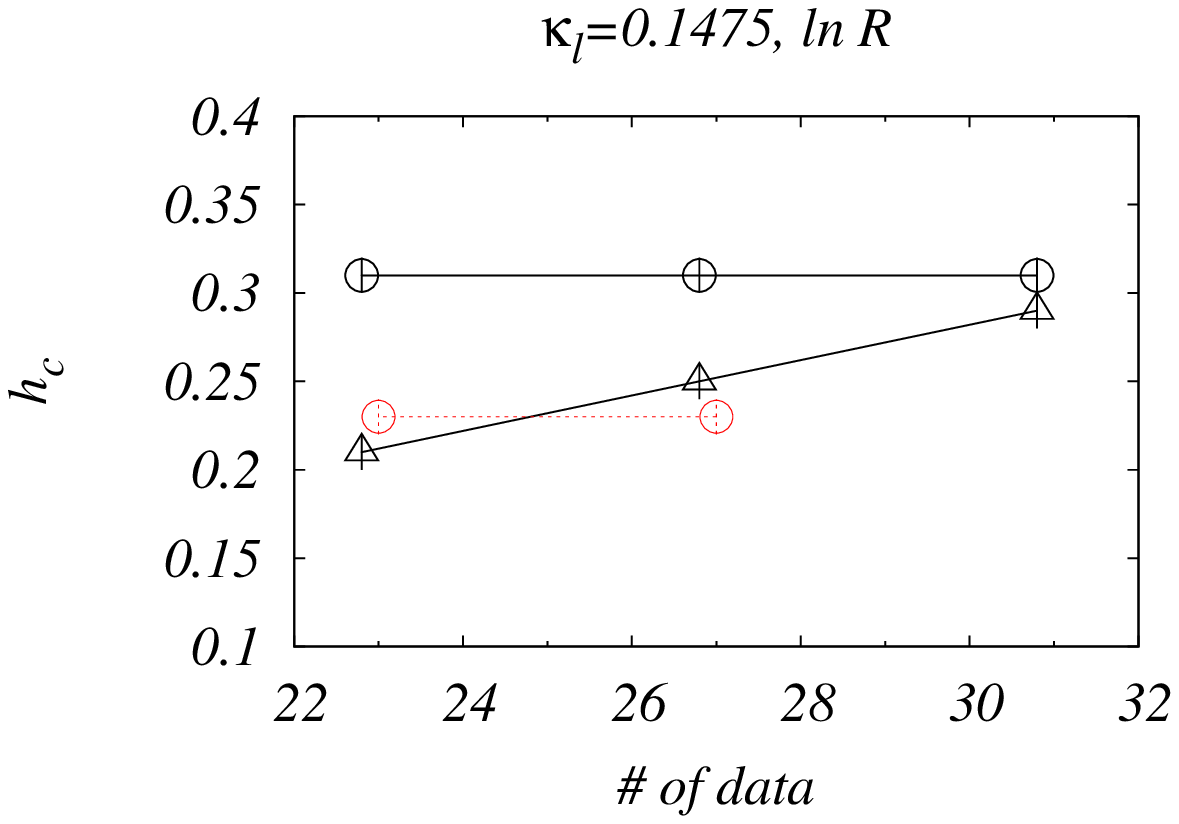}\\[-2ex]
\includegraphics*[width=0.5 \textwidth,clip=true]
{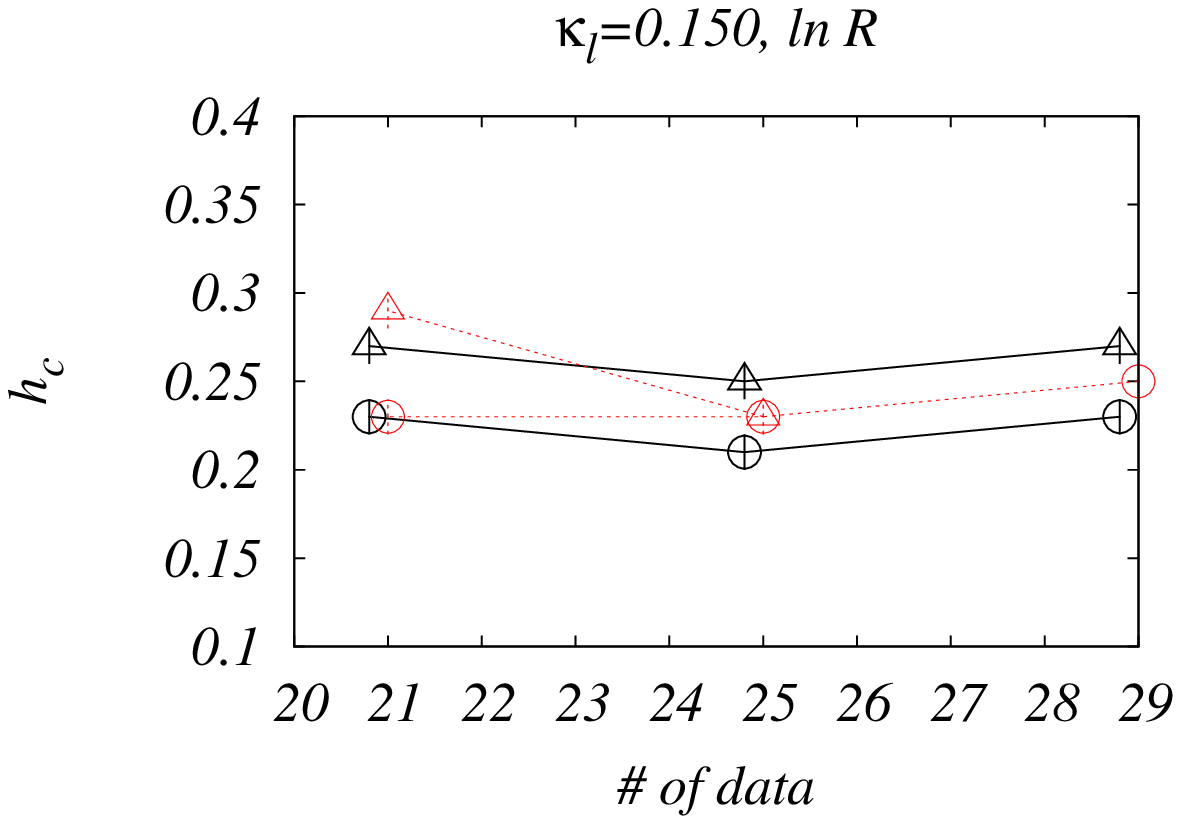}&
\includegraphics*[width=0.5 \textwidth,clip=true]
{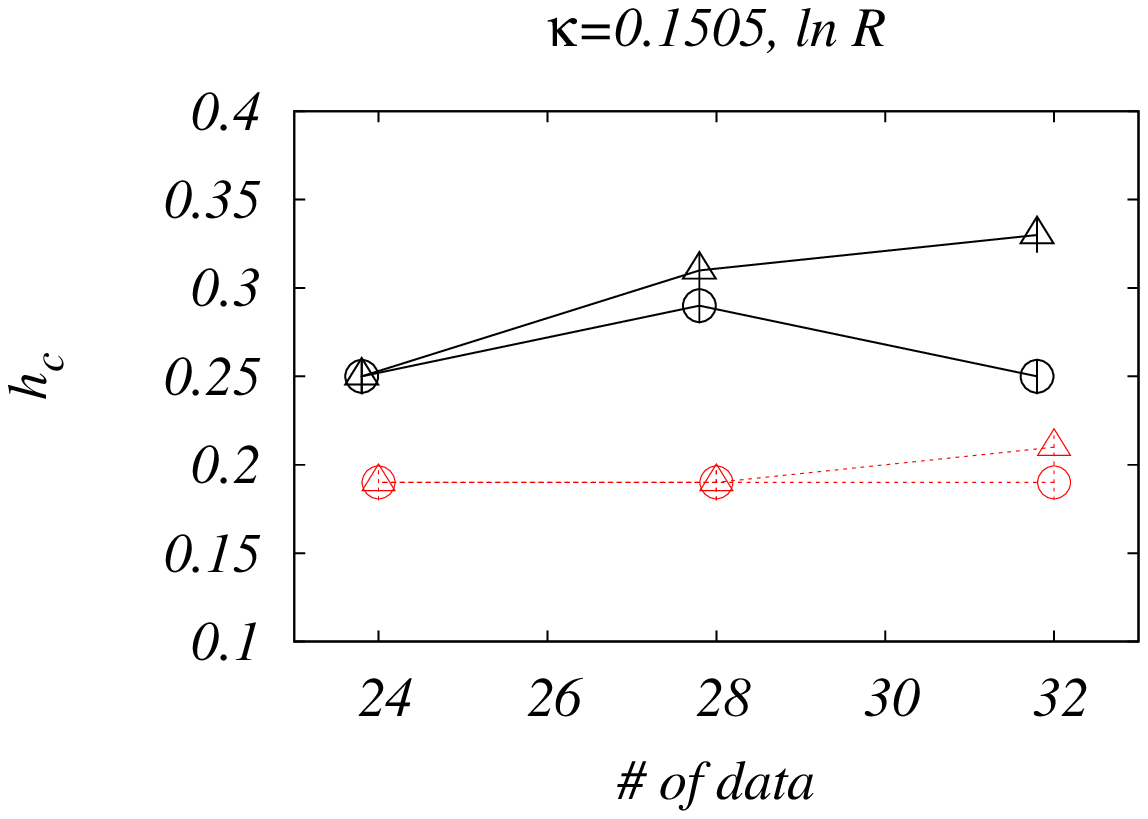}\\[-2ex]
\includegraphics*[width=0.5 \textwidth,clip=true]
{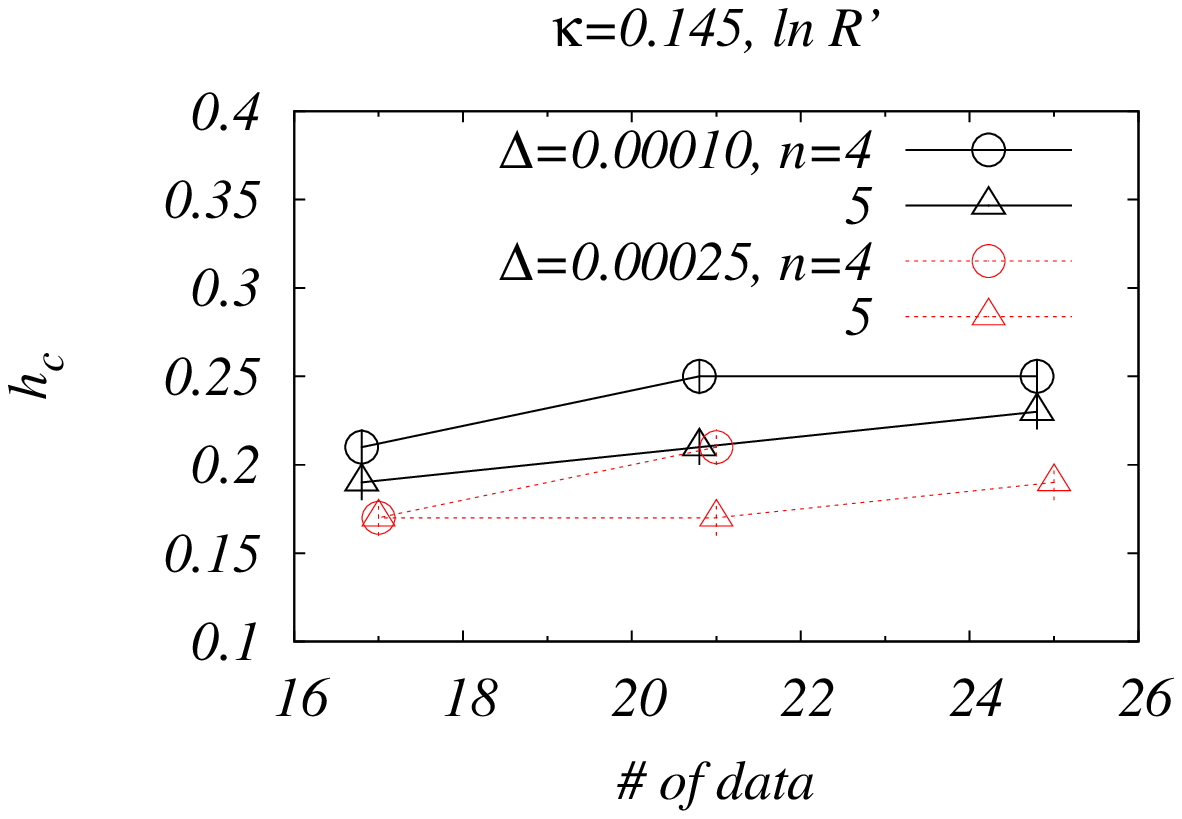}&
\includegraphics*[width=0.5 \textwidth,clip=true]
{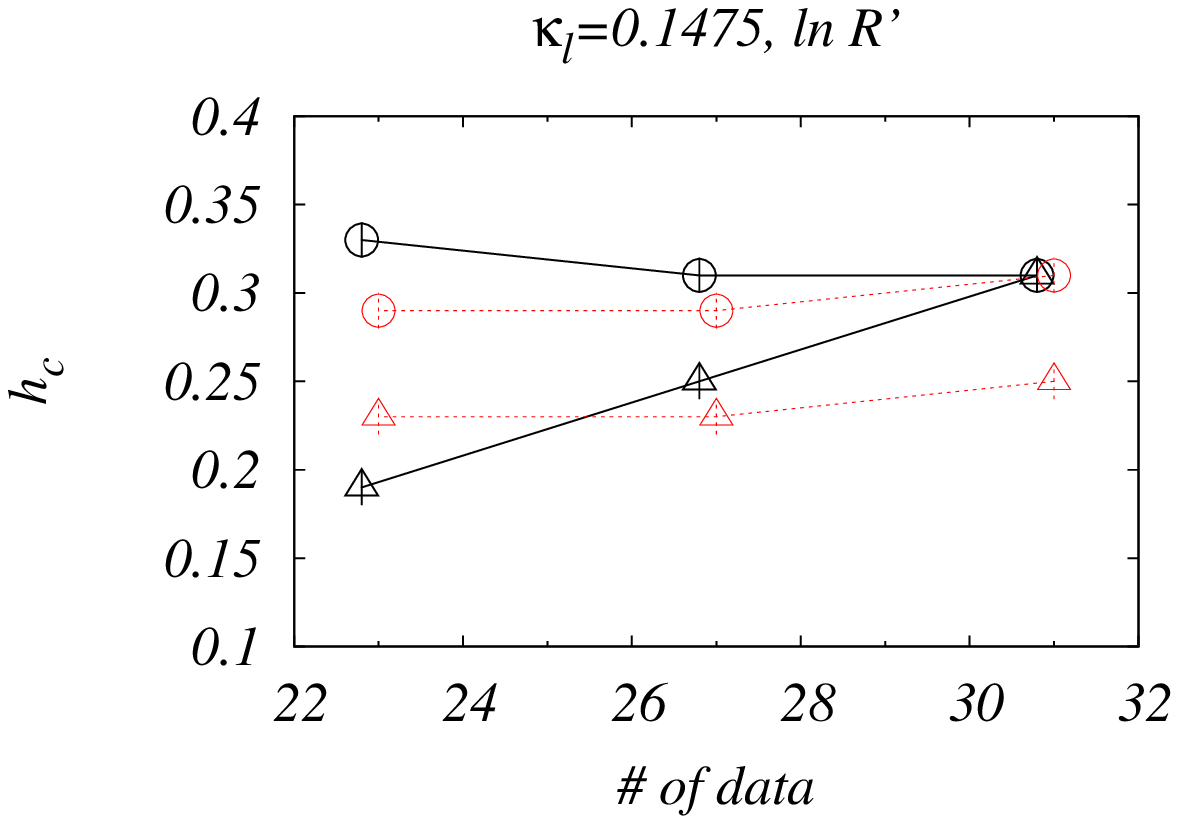}\\[-2ex]
\includegraphics*[width=0.5 \textwidth,clip=true]
{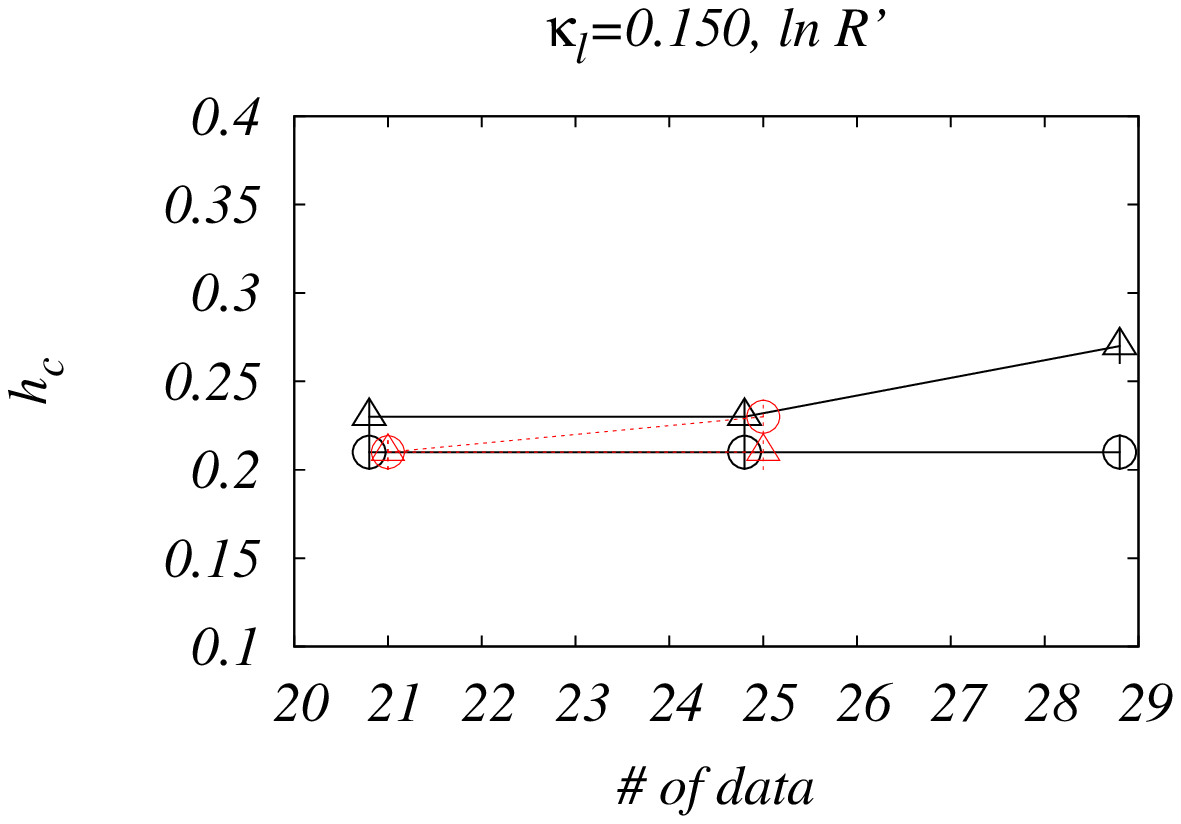}&
\includegraphics*[width=0.5 \textwidth,clip=true]
{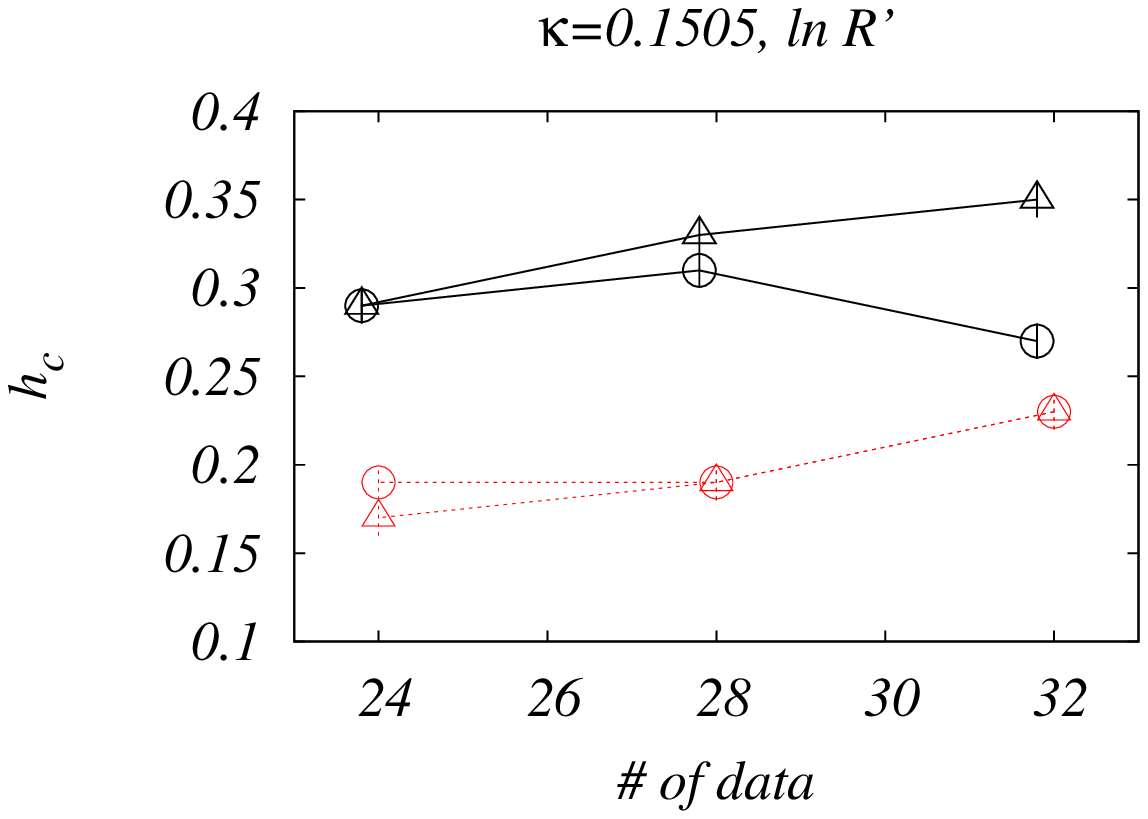}\\[-2ex]
\end{tabular}
\vspace{-2ex}
 \caption{The critical value of $h$ as a function of the number of the
 data points used.
 }
\label{fig:scan-hc}
\end{center}
\end{figure}

To determine $h_c$, we iterate the calculation with $h$ varying in
steps of 0.02.
Figure~\ref{fig:scan-hc} shows the critical values of $h$
as a function of the number of data points used in the fit,
corresponding to Fig.~\ref{fig:chi2}.
Since no reason exists to select the best result from them,
we take all the results satisfying $\chi^2/$dof $<$ 3 as the final
results, and the systematic uncertainty is chosen to cover the whole
accepted results.

\begin{figure}[tb]
\vspace*{-1ex}
\begin{center}
\begin{tabular}{c}
\includegraphics*[width=0.6 \textwidth,clip=true]
{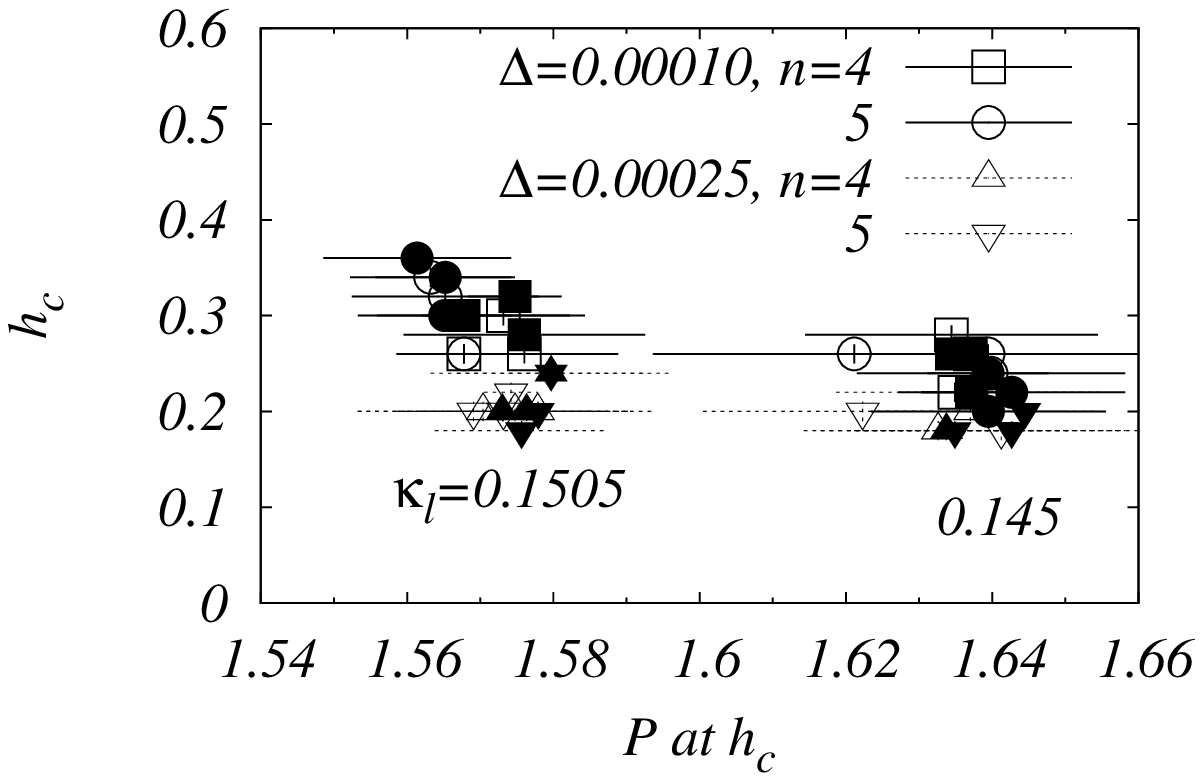}\\
\end{tabular}
\vspace{-1ex}
\caption{$h_c$ and $P$ at $h_c$: The results from $\ln R$ (open symbols)
 and $\ln R'$ (filled symbols).}
\label{fig:hc-P}
\end{center}
\vspace{-2ex}
\end{figure}

We can also determine the value of $P$ at $h_c$, denoted by $P_c$
(for the numerical values, see Table~\ref{tab:hc-result}).
In Fig.~\ref{fig:hc-P}, all the results with $\chi^2/{\rm dof} < 3$ are
plotted together on the $P_c$-$h_c$ plane.
While $h_c$ is insensitive to the two-flavor mass, $P_c$ is found
to decrease towards the chiral limit of the two-flavor mass.
This qualitative feature is tested in the direct simulations of
$2+\Nf$-flavor QCD in Sec.~\ref{subsec:direct-sim}.

\begin{table}[tb]
 \centering
 \begin{tabular}{c|cc}
 $\kl$ & $h_c$ & $P$ at $h_c$\\
  \hline
   0.1450 & 0.23( 6) & 1.627(34) \\
   0.1475 & 0.27( 8) & 1.595(27) \\
   0.1500 & 0.26( 5) & 1.561(41) \\
   0.1505 & 0.27(10) & 1.572(24) \\
 \end{tabular}
 \caption{Numerical results of $h_c$ and $P$ at $h_c$.
 }
 \label{tab:hc-result}
\vspace{-1ex}
\end{table}

\begin{figure}[t]
\vspace*{-1ex}
\begin{center}
\begin{tabular}{cc}
\includegraphics*[width=0.5 \textwidth,clip=true]
{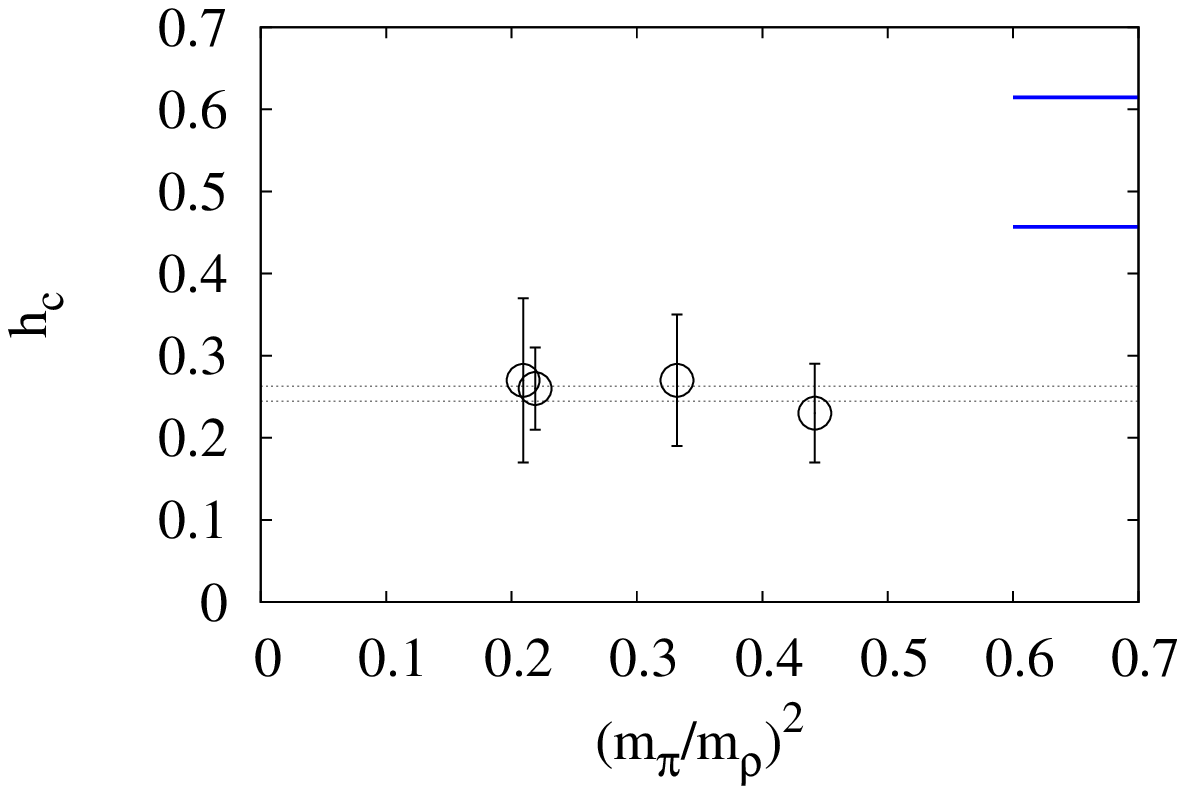} &
\includegraphics*[width=0.5 \textwidth,clip=true]
{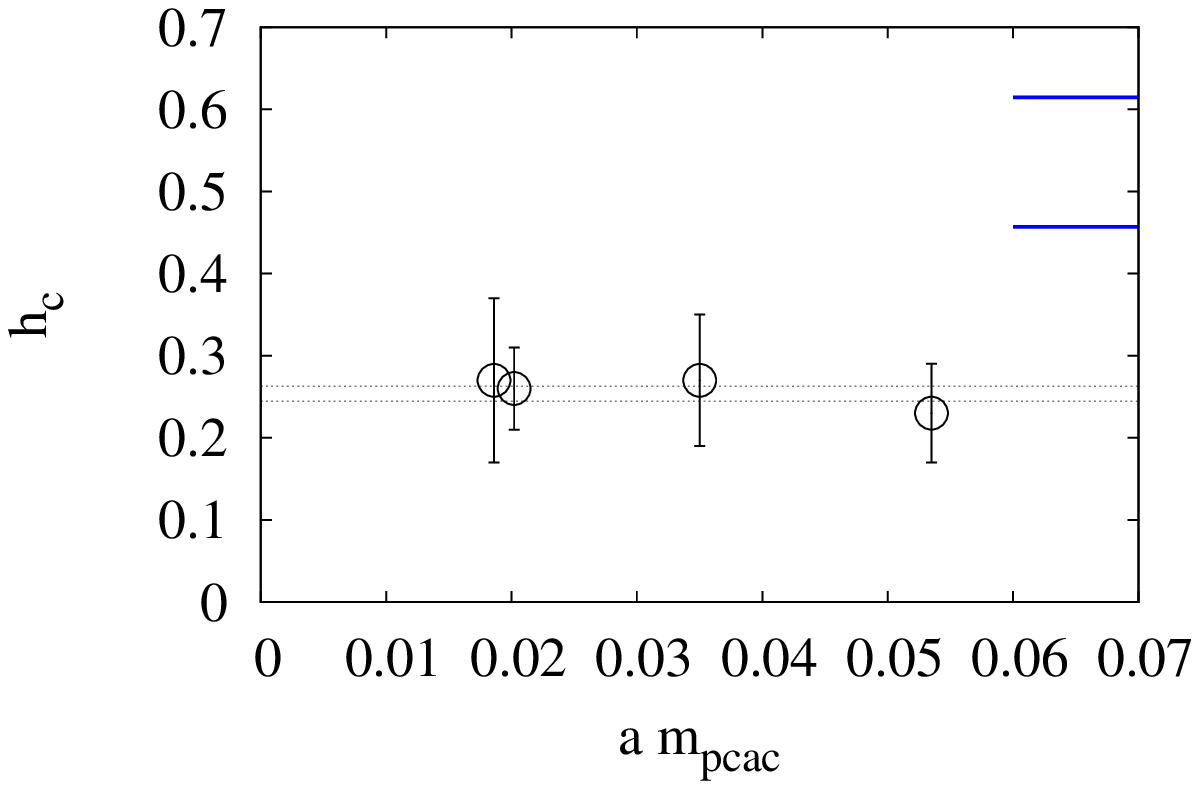} \\
\includegraphics*[width=0.5 \textwidth,clip=true]
{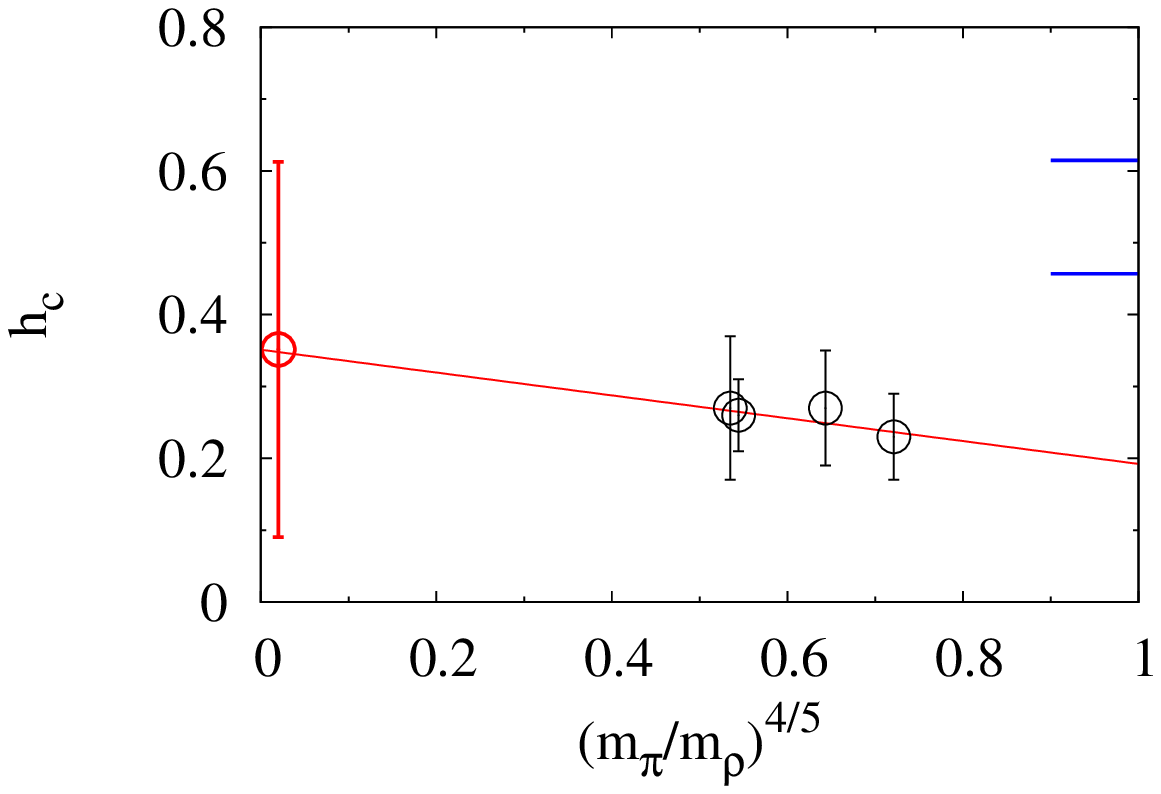} &
\includegraphics*[width=0.5 \textwidth,clip=true]
{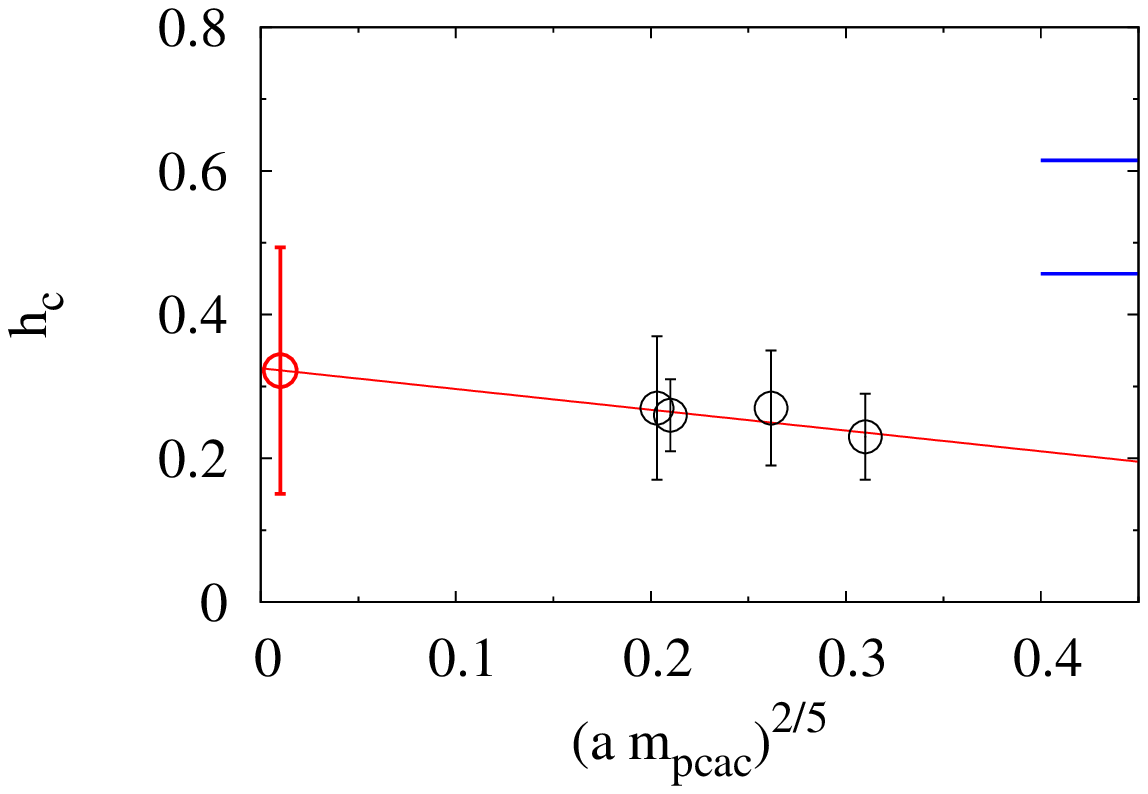} \\
\end{tabular}
\vspace{-2ex}
\caption{The light quark mass dependence of $h_c$.
 Top: A constant fit.
 Bottom: A linear fit is performed with the horizontal axis
 proportional to $\ml^{2/5}$, testing the mean field prediction
 Eq.~(\ref{eq:meanfield-scaling}).
 The band appearing in the top right corner is $h_c$ determined in the
 direct simulation with $\kappa_l=0$ and $\Nf=50$ (for details, see
 Sec.~\ref{subsec:direct-sim}).
 }
\label{fig:lq-dep}
\vspace{-2ex}
\end{center}
\end{figure}

The light quark mass dependence of $h_c$ is plotted in
Fig.~\ref{fig:lq-dep} as a function of the $m_\pi/m_\rho$ ratio (left)
and the PCAC quark mass (right).
The error is dominated by the systematic uncertainty associated with
the fitting procedure.
In the top-right corner of each plot, $h_c$ determined from the direct
$2+\Nf$ flavor simulation with $\kl=0$ ({\it i.e.} $\ml=\infty$) and
$\Nf=50$ is shown with an uncertainty (for the direct simulations, see
Sec.~\ref{subsec:direct-sim}).
$h_c$ at $\kl=0$ is clearly larger than those around
$0.145 \le \kl \le 0.1505$, which indicates that $h_c$ gradually
decreases towards the chiral limit as a global behavior.
In the range of $0.145 \le \kl \le 0.1505$, $h_c$ does not show
significant dependence on $\ml$ within the error, and a constant fit
yields $h_c=0.23(1)$ in the chiral limit.

If this mild dependence is continued down to the chiral limit and hence
$h_c$ in the chiral limit remains positive and finite, $h_c$ in the
chiral limit corresponds to the tricritical point in
Fig.~\ref{fig:columbia-plot}.
Mean field analysis of an effective theory predicts the tricritical
scaling~\cite{Wilczek:1992sf,Rajagopal:1992qz,Ukawa:1995tc,Ejiri:2008nv},
\begin{eqnarray}
 h_c \sim (\mbox{const.})\times \ml^{2/5} + \mbox{const.}
 \label{eq:meanfield-scaling}
\end{eqnarray}
in the vicinity of the tricritical point, where the power $2/5$ is
independent of $\Nf$.
In addition to the constant fit (top), we also fit the data to a linear
function of $\ml^{2/5}$ in each plot, yielding $h_c=0.35(26)$ and
$0.32(17)$ in the chiral limit of the two-flavor mass, respectively.
Note that the slope is undetermined and consistent with zero.
In either case, a positive value of $h_c$ is favored in the chiral
limit, which suggests the second order transition of massless two-flavor
QCD.
Further checks require more extensive lattice calculations and are
postponed to future papers.

\section{Consistency check}
\label{sec:consistency-check}

\subsection{Effective potential constraining Polyakov loop}
\label{subsec:polyakov}

As an independent check of the results obtained in
Sec.~\ref{sec:numerical_results}, we try to estimate $h_c$ with a
different method.
The quantity to be constrained to obtain the PDF is arbitrary as long as
it has an overlap with the order parameter.
In this section, we take the real part of the Polyakov loop, $\hat L$.
Recalling Eq.~(\ref{eq:effective-potential-2}), the constraint effective
potential for $\hat L$ is given by
\begin{eqnarray}
      V_L(L;\beta_{\rm ref},\kl,\kh,\Nf)
&=& - \ln w_L(L;\beta_{\rm ref},\kl,0,0)
    - \ln \left(\frac{w_L(L;\beta_{\rm ref},\kl,\kh,\Nf)}
                     {w_L(L;\beta_{\rm ref},\kl,0  ,0)}
		   \right)\\
&=&   V_{L,\,\rm light}(L; \beta_{\rm ref},\kl)
    - \ln R_L(L;\beta_{\rm ref},\kl,\kh,\Nf)\ ,
 \label{eq:effective-potential-poly}
\end{eqnarray}
where
\begin{eqnarray}
      V_{L,\,\rm light}(L; \beta_{\rm ref},\kl)
&=& - \ln w_L(L;\beta_{\rm ref},\kl,0,0) \\
&=& - \ln \langle\,
             \delta(L-\hat L)\,
             e^{6\,(\beta_{\rm ref}-\beta)\,N_{\rm site}\, \hat P}\,
          \rangle_{(\beta,\kl)}\, ,\\
    R_L(L; \beta_{\rm ref},\kl,\kh,\Nf)
&=& e^{6\,N_s^3\,h\,L}
    \left\langle \displaystyle
       \exp\left[
            36\,N_s^3\,h\,\hat W_P
	  +  6\,(\beta_{\rm ref}-\beta)\,N_{\rm site}\, \hat P
	   \right]
    \right\rangle_{L: {\rm fixed},(\beta,\kl)}\ .
\end{eqnarray}
Unlike the case with $\hat X=\hat P$, $\beta_{\rm ref}$ dependence
remains in the second derivative of the potential with respect to $L$,
which means $\beta_{\rm ref}$ has to be explicitly tuned to the
(pseudo)critical temperature for each value of $h$.
In the crossover region, the minimum of the potential and the minimum of
its second derivative are realized at the same value of $L$.
On the other hand, in the first order region, $\beta_{\rm ref}$ has to
be tuned until the two minima in the potential take the same depth.

\begin{figure}[tb]
\begin{center}
\begin{tabular}{cc}
\includegraphics*[width=0.5 \textwidth,clip=true]
{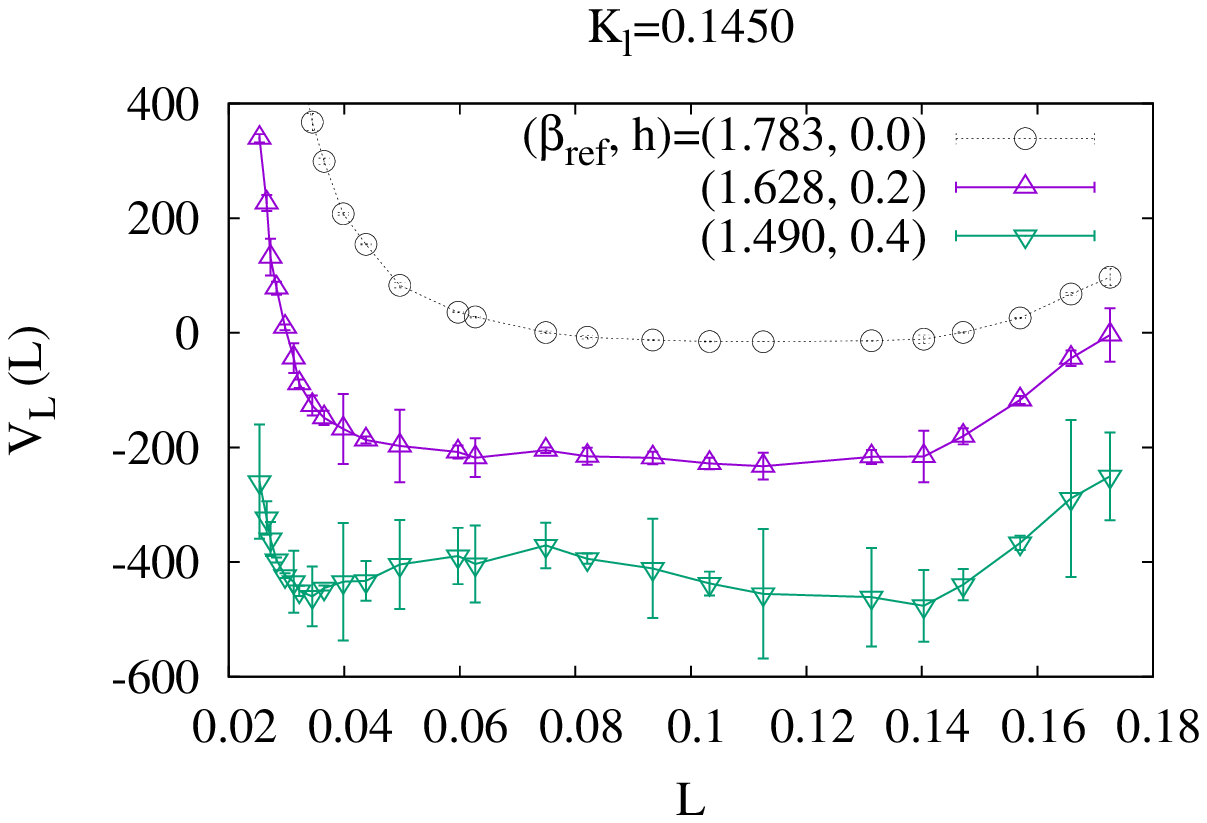}&
\includegraphics*[width=0.5 \textwidth,clip=true]
{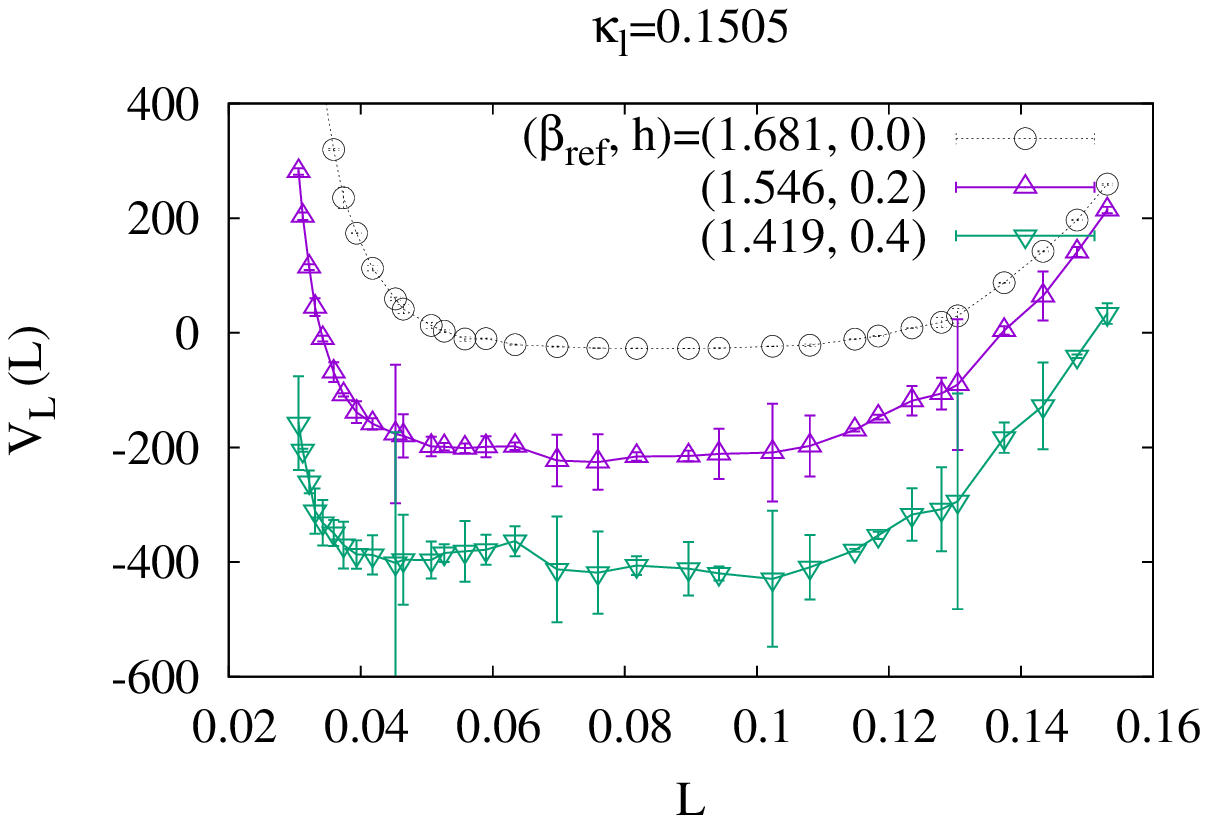}
\end{tabular}
\vspace{-1ex}
\caption{The effective potential constructed by constraining $\hat L$
 near the critical temperatures.
 The absolute values of the potential are shifted for comparison.
 The results at $h=0.0$, 0.2 and 0.4 are shown for $\kl=0.145$ (left)
 and 0.1505 (right).
 }
\label{fig:Veff-RePoly}
\end{center}
\end{figure}

The potential, $V_L(L)$, is shown in Fig.~\ref{fig:Veff-RePoly}
for $h=0.0$, 0.2 and 0.4, where $V_L(L)$ at each $h$ is shifted in a
vertical direction for comparison.
First of all, the effective potential for the Polyakov loop is not as
clean as that for the generalized plaquette, especially at nonzero $h$,
which only allows us to extract the qualitative feature.
It is seen that the potential has a positive curvature at $h=0$ in our
lightest and heaviest light quarks.
When $h=0.2$, the potential around the minimum becomes almost flat,
indicating that it is close to the end point.
If $h$ is further increased to 0.4, the double well shape appears to emerge
though the statistical error makes it ambiguous.
These qualitative features are consistent with the findings in the
previous section.

\subsection{Direct simulations of $2+\Nf$ flavor QCD}
\label{subsec:direct-sim}

Although the convergence of hopping parameter expansion is not the
matter for the discussion in the previous section, it is interesting to
investigate the convergence of the HPE for future applications.
We study this by explicitly performing simulations of $2+\Nf$ flavor QCD
and comparing the results with those based on the HPE.
However, thoroughly precise calculations of the many flavor system
require an extensive scan of simulation parameters ($\kh$, $\beta$ and
$\Nf$) even after fixing $\kl$.
Furthermore, in general, it is not easy to locate the end point of the
the first order transition accurately, because the statistical noise
grows as one approaches the end point.
Instead, we draw a thermal cycle on the $\beta$-$P$ plane, which is
obtained by increasing (or decreasing) $\beta$ until passing its
(pseudo) critical value and then reversing the direction.
For representative values of $\kh$, $\kl$ and $\Nf$, we accumulated
400 trajectories at each point of the thermal cycle.

\begin{figure}[tb]
\begin{center}
\begin{tabular}{cc}
\includegraphics*[width=0.5 \textwidth,clip=true]
{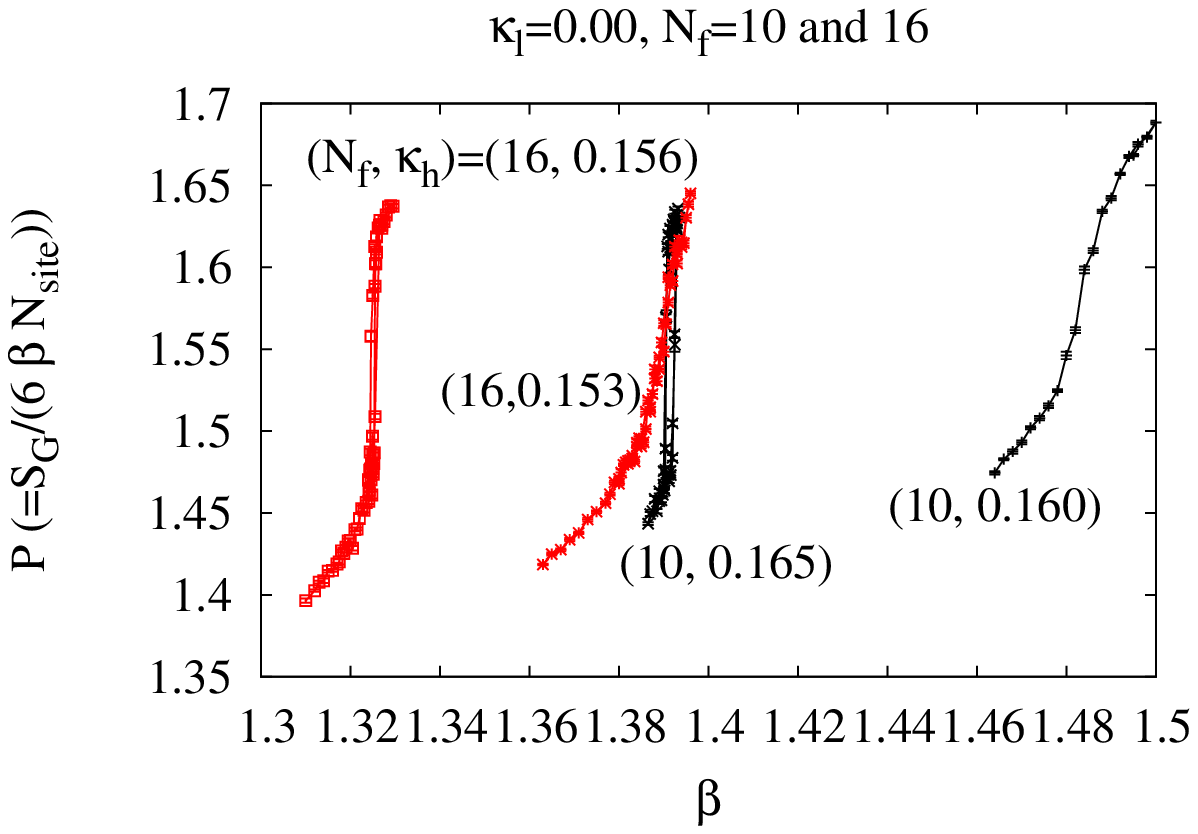}&
\includegraphics*[width=0.5 \textwidth,clip=true]
{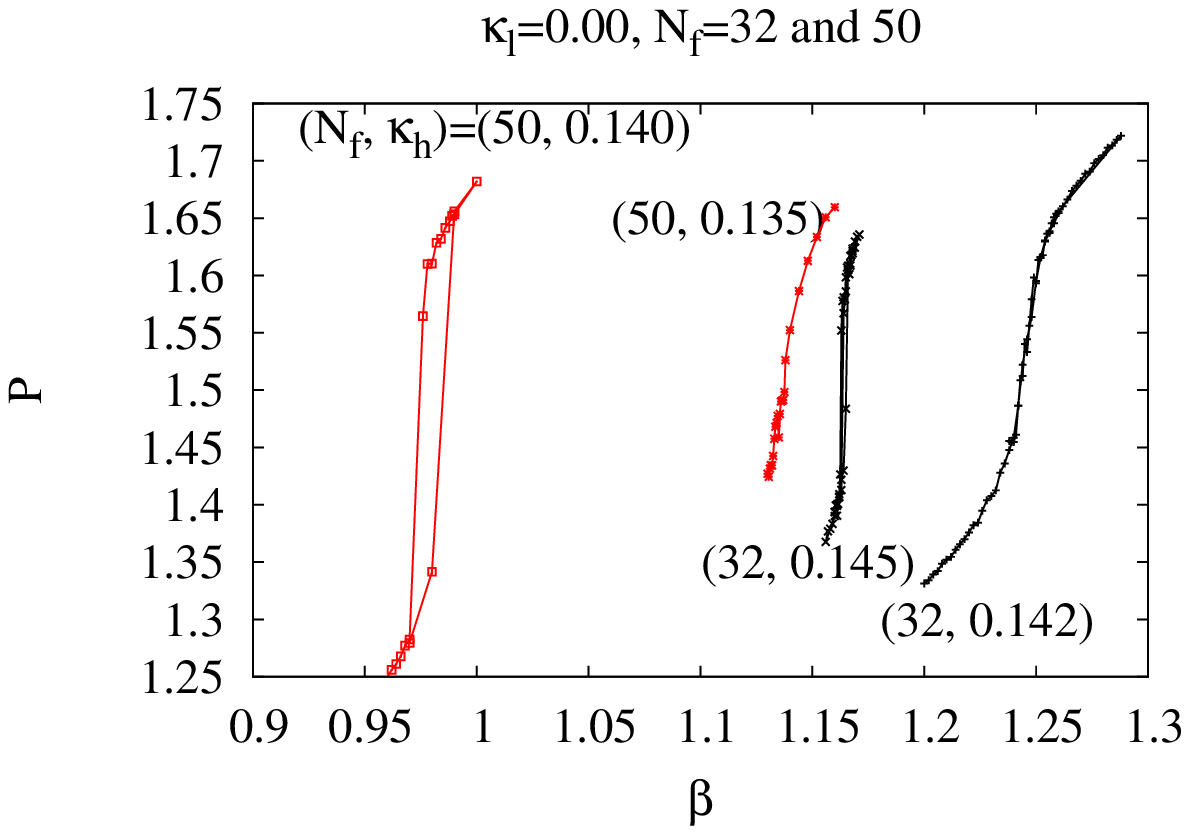}\\[-2ex]
 (a)&(b)\\[2ex]
\includegraphics*[width=0.5 \textwidth,clip=true]
{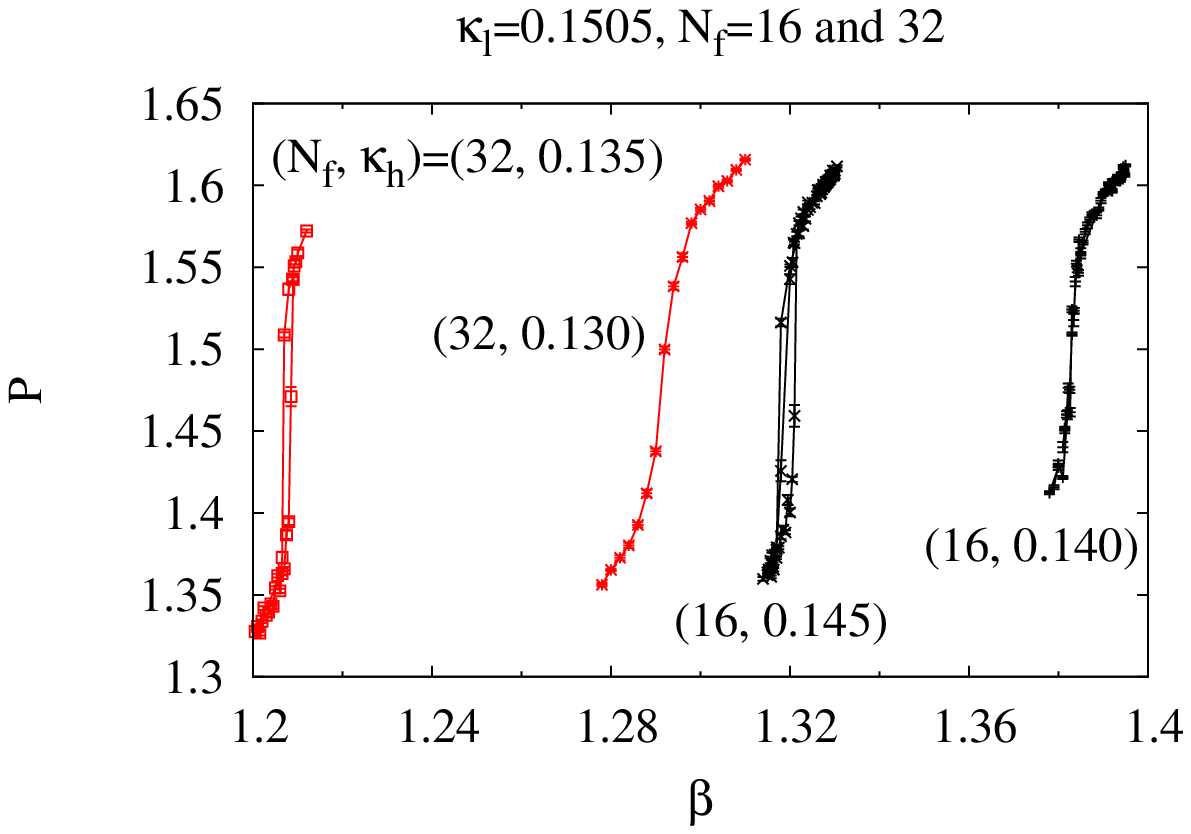}&
\includegraphics*[width=0.5 \textwidth,clip=true]
{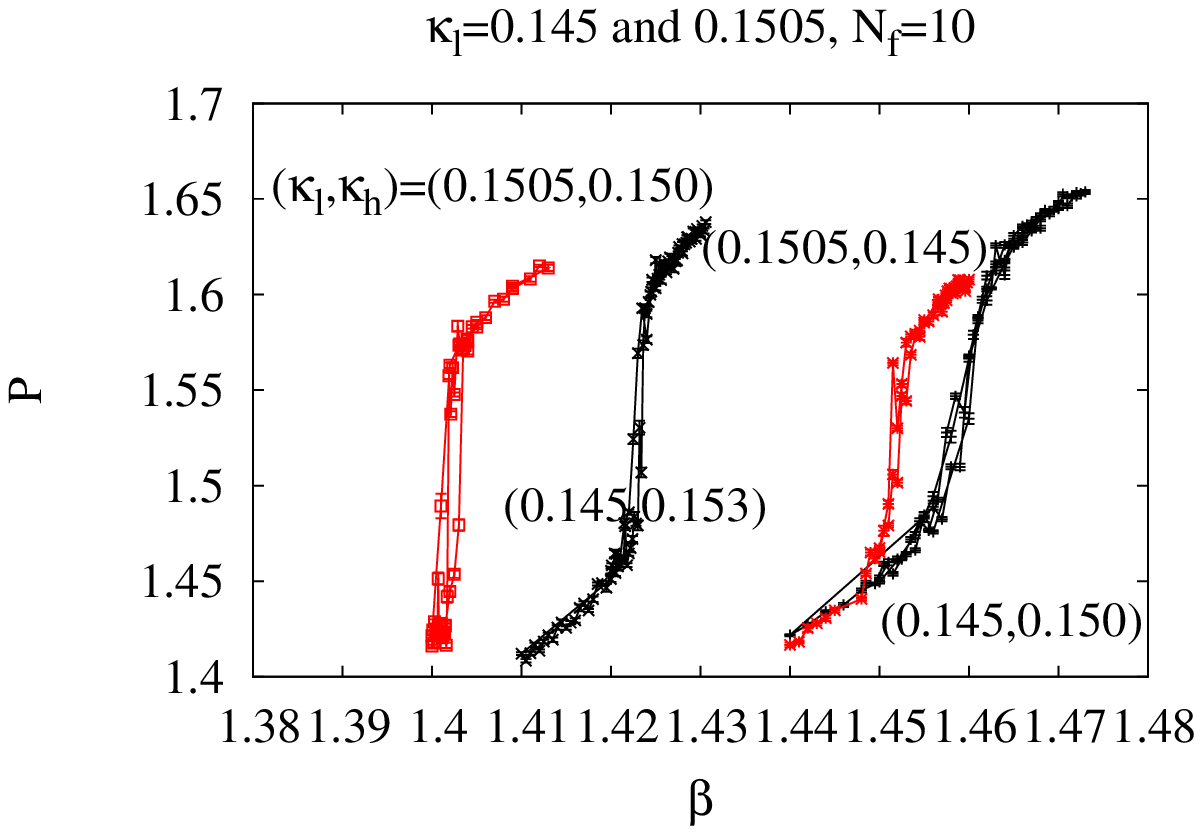}\\[-2ex]
 (c)&(d)\\[2ex]
\end{tabular}
\vspace{-1ex}
\caption{The thermal cycle on the ($P$, $\beta$)-plane in $2+\Nf$ flavor
 QCD. The vicinity of the end point, {\it i.e.} $\kh\approx\khc$, is
 shown.}
\label{fig:hysteresis}
\end{center}
\end{figure}

We take $\kl=0, 0.145$ and 0.1505, and choose $\Nf$ ranging from 4 to
50, depending on $\kl$, and monitor each thermal cycle whether the
hysteresis curve occurs or not.
If it occurs, the parameters chosen turn out to be in the first order
region.
By repeating this, we try to find the critical value of $\kh$, $\khc$,
for fixed values of $\kl$ and $\Nf$.
The thermal cycles at several simulation parameters are shown in
Fig.~\ref{fig:hysteresis}.

Figures~\ref{fig:hysteresis} (a) and \ref{fig:hysteresis} (b) show the
thermal cycle at $\kl=0$ and $\Nf=10$, 16, 32, 50, which tells us that
$\khc$ decreases with $\Nf$.
Figure~\ref{fig:hysteresis} (c) shows the same but with nonzero $\kl$.
Comparing the $\Nf=16$ data in Figs.~\ref{fig:hysteresis} (a) and
\ref{fig:hysteresis} (c) or the $\Nf=32$ data in
Figs.~\ref{fig:hysteresis} (b) and \ref{fig:hysteresis} (c), it is found
that $\khc$ is clearly different between $\kl=0$ and 0.1505.
In Fig.~\ref{fig:hc-P} of Sec.~\ref{subsec:lat-para}, we have discussed
that the HPE predicts that $P$ at $h_c$ decreases with $\kl$.
Figure~\ref{fig:hysteresis} (d) seems to show that it is the case, at
least, qualitatively.

\begin{figure}[tb]
\begin{center}
\begin{tabular}{cc}
\includegraphics*[width=0.7 \textwidth,clip=true]
{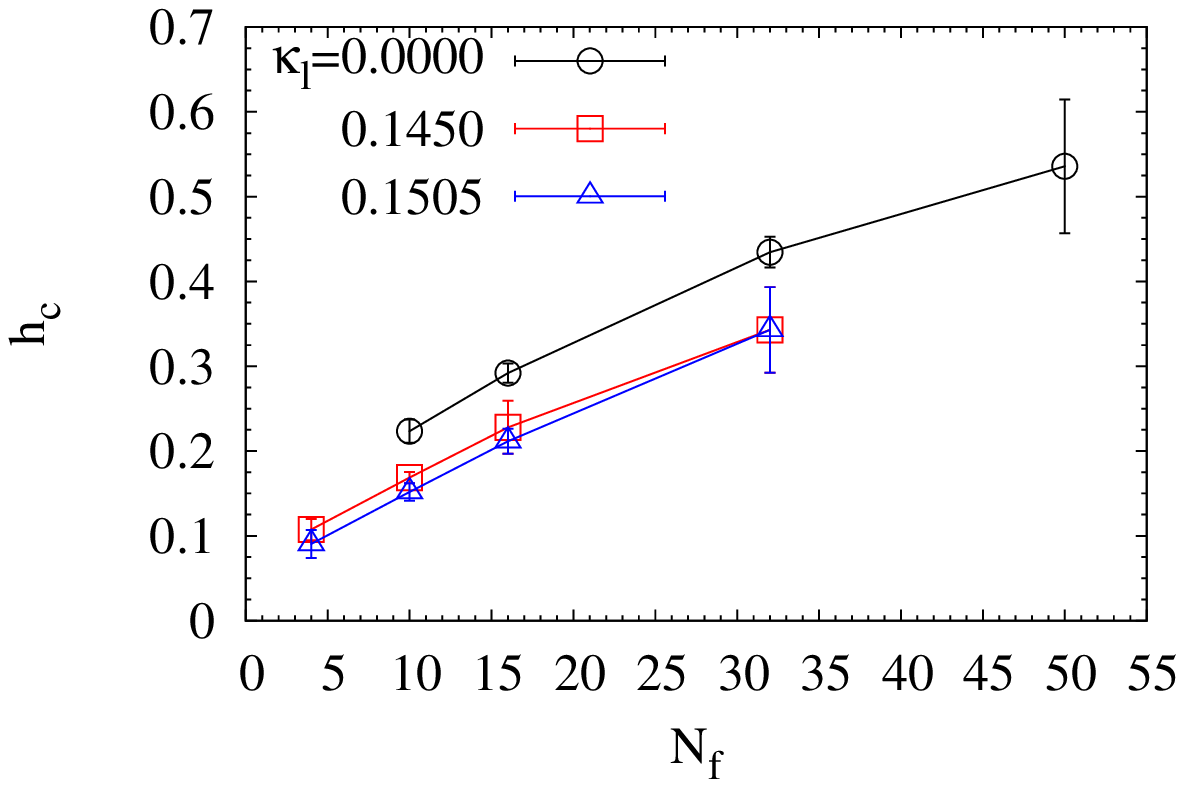}\\
\end{tabular}
\vspace{-1ex}
\caption{$\Nf$ dependence of $h_c$ through direct simulations of $2+\Nf$
 flavor QCD.}
\label{fig:direct-hc}
\end{center}
\end{figure}

The critical values, $\kappa_{h_c}$, obtained at each $\kl$ and $\Nf$
are translated into $h_c$ using Eq.~(\ref{eq:h}), and plotted as a
function of $\Nf$ in Fig.~\ref{fig:direct-hc}.
It is expected that, for a sufficiently large $\Nf$,
$h_c$ approaches an asymptotic value, but the value of $\Nf$ we have
studied seems not to reach such a region yet although the increasing
rate looks slowing down.
Importantly, it is seen that $\khc$ does not differ by much between
$\kl=0.145$ and 0.1505, which is consistent with the observation in the
HPE analysis.
It is interesting to include the next-to-leading order contribution of
the HPE analysis.

\section{Summary and discussion}
\label{sec:summary}

We have studied the finite temperature phase transition of QCD with two
light and many heavy quarks at zero chemical potential, where the heavy
quarks are introduced in the form of the hopping parameter expansion via
the reweighting method.
The phase structure was scanned on the $\kl$-$h$ plane to identify the
critical line separating the continuous crossover and the first order
regions.

The nature of the transition is identified by the shape of the
constraint effective potential constructed from the probability
distribution function of the generalized plaquette.
For $h=0$, the system reduces to two-flavor QCD, which always shows
continuous crossover and hence the potential has a single well.
As one increases $h$, at some point the potential takes a double-well
shape, which defines a critical value, $h_c$.
We have determined $h_c$ at four light quark masses, and observed that
$h_c$ is independent of two-flavor mass in the range we have studied
($0.46 \le m_\pi/m_\rho\le 0.66$).
This result indicates that the critical heavy mass remains finite in the
chiral limit of the two flavors, suggesting the phase
transition of massless two-flavor QCD is of second order.
Some of the qualitative features observed in the main analysis were
checked by two independent analyses.

The approach in this study can be said as follows.
Two-flavor QCD with a finite mass is enforced to undergo a first order
phase transition by adding extra quarks.
It is then likely that those extra quarks are necessary to keep the
first order transition down to the chiral limit of two-flavor QCD.
This method is applicable for any kinds of lattice fermions unless they
contain $\beta$ dependent coefficients.
According to the definition of $h$, Eq.~(\ref{eq:h}), $\khc$ can be
considered to be arbitrarily small for a given $h_c$ by assuming
arbitrarily large $\Nf$, and thus we do not have to care about the
convergence of the hopping parameter expansion.
Nevertheless, it is interesting to see the limitation of the HPE for a
fixed $\Nf$ for further applications.

In order to establish our finding, possible systematic uncertainties,
which are not investigated in the present paper, need to be
understood.
Since, at the end point, the second order phase transition occurs, a
sizable finite volume effect is possible in the vicinity of the point.
However, we do not expect it to be significant in the many flavor
approach.
In this approach, the end point is determined through the extrapolation
of the effective potential with regard to $h$, and the extrapolation is
performed in a region free from finite size effects since the two-flavor
configurations are all generated at a parameter region away from the
second order end point.
Namely, at the price of the uncertainty due to the extrapolation, we
could have avoided the finite volume effect associated with the second
order phase transition.
Nevertheless, it is clearly important to explicitly check that the
effect is under good control.
Such work is ongoing.

Although the behavior observed in Fig.~\ref{fig:lq-dep} seems to suggest
that the chiral limit of $h_c$ is finite and positive, it then has to
show the tricritical scaling [Eq.~(\ref{eq:meanfield-scaling})].
At the present, our results allow us to fit to any smooth function
in $\ml$.
In order to improve the situation, we need to explore lighter quark
masses and reduce the systematic uncertainty associated with the
fitting procedure.
However, during the preliminary study, we realized that the lightest
quark mass presented in this paper is the lower limit in our lattice
setup.
To go beyond the limit, the setup has to be changed.
Towards the ultimate goal, the discretization effects also have to be
examined.
A systematic study of these uncertainties requires large scale
simulations, and we postpone them to future works.

We can extend the many flavor approach to explore QCD at finite chemical
potential as initiated in Refs.~\cite{Ejiri:2012rr,Iwami:2015mqa}.
In this case, mean field analysis predicts that the critical line
runs like ${\ml}^c \sim |\mu|^5$~\cite{Ejiri:2008nv}.
We believe that such a study brings valuable information to
understand the rich QCD phase diagram.

\section*{Acknowledgments}

We would like to thank members of the WHOT-QCD Collaboration for
useful discussions.
We also thank Ken-Ichi Ishikawa
for providing us his simulation codes.
This work is in part supported by
JSPS KAKENHI Grant-in-Aid for Scientific Research (B)
(No.\ 15H03669 [N.~Y.],
       26287040 [S.~E.]
)
and (C)
(No.\ 26400244 [S.~E.]), and by the Large Scale Simulation Program of High
Energy Accelerator Research Organization (KEK) No.\ 14/15-23.

\end{document}